\UseRawInputEncoding
\documentclass{SCIS2023}

 \usepackage[numbers,sort&compress]{natbib}

\begin{document}
\ArticleType{REVIEW}
\Year{2023}
\Month{}
\Vol{}
\No{}
\DOI{}
\ArtNo{}
\ReceiveDate{}
\ReviseDate{}
\AcceptDate{}
\OnlineDate{}

\title{Nonadiabatic Holonomic Quantum Computation and Its Optimal Control}

\author[1]{Yan Liang}{}
\author[1]{Pu Shen}{}
\author[1,2]{Tao Chen}{}
\author[1,2,3]{Zheng-Yuan Xue}{{zyxue83@163.com}}

\AuthorMark{Liang Y, Shen P, Tao Chen, Xue Z-Y}

\AuthorCitation{Liang Y, Shen P, Tao Chen, Xue Z-Y}


\address[1]{Key Laboratory of Atomic and Subatomic Structure and Quantum Control (Ministry of Education),\\ and School of Physics, South China Normal University, Guangzhou  {\rm 510006}, China}
\address[2]{Guangdong Provincial Key Laboratory of Quantum Engineering and Quantum Materials, \\  Guangdong-Hong Kong Joint Laboratory of Quantum Matter,  and Frontier Research Institute for Physics,\\  South China Normal University, Guangzhou {\rm 510006}, China}
\address[3]{Hefei National Laboratory,  Hefei 230088, China}

\abstract{Geometric phase has the intrinsic property of being resistant to some types of local noises as it only depends on global properties of the evolution path. Meanwhile, the non-Abelian geometric phase is in the matrix form, and thus can naturally be used to implement high performance quantum gates, i.e., the so-called holonomic quantum computation. This article reviews recent advances in nonadiabatic holonomic quantum computation, and focuses on various optimal control  approaches that can improve the gate performance, in terms of the gate fidelity and robustness. Besides, we also pay special attention to its possible physical realizations and some concrete examples of experimental realizations. Finally, with all these efforts, within state-of-the-art technology, the performance of the implemented holonomic quantum gates can outperform the conventional dynamical ones, under certain conditions.}

\keywords{Quantum computation, geometric phases,  quantum gates, optimal control}

\maketitle

\section{Introduction}

Quantum computation has the potential ability to solve certain hard problems, such as large integer factorization \cite{shor1994proceedings} and fast searching \cite{grover1997quantum}, in a way that is much faster than any known classical computers. The practical application of quantum computation requires precise quantum gates, in terms of fidelity and robustness, which is challenging with current technology, as the needed quantum evolution will be inevitably disturbed due to various control errors and the environment-induced decoherence effect. It is well-known that geometric phase only depends on the global properties of the evolution path and is independent of the evolution speed. Therefore, quantum computation based on geometric phases \cite{berry1984quantal,wilczek1984appearance,aharonov1987phase,anandan1988non}, i.e., geometric quantum computation \cite{zanardi1999holonomic,pachos1999non}, has become one of the inherently robust fault-tolerant quantum computation schemes. To date, geometric quantum computation is proposed based on cyclic/non-cyclic Abelian/non-Abelian  adiabatic/nonadiabatic geometric phases.

Here, we term the quantum computation based on the non-Abelian geometric phase as holonomic quantum computation (HQC) \cite{pachos1999non,2021Research,zhang2021geometric}. As detailed in Section \ref{sectionAHQC}, early HQC proposals were based on the adiabatic evolution of quantum systems with degenerate eigenstates, which is restricted by the cyclic evolution condition.  The physical implementation of HQC has been proposed in many quantum systems, such as trapped ions \cite{duan2001geometric}, atoms \cite{recati2002holonomic}, superconducting quantum circuits \cite{siewert2002holonomic,faoro2003non}, and semiconductor quantum dots \cite{solinas2003holonomic}, etc. However, implementing adiabatic HQC is difficult, and achieving high quality quantum gates with this method seems challenging in typical solid-state quantum systems. Firstly, the evolution process needs to be slow enough to meet the adiabatic condition \cite{tong2010quantitative}, where decoherence will  lead to considerable error. Secondly, manipulation of the complex energy level structure is required for HQC, which is experimentally demanding. Thus, only recently, the elementally adiabatic holonomic quantum gates have been demonstrated with trapped ions \cite{toyoda2013realization} and cold atomic gas \cite{leroux2018non}.

As presented in Section \ref{NHQC}, the main theoretical obstacle of slow evolution can be resolved by using nonadiabatic evolution, i.e., the nonadiabatic holonomic quantum computation (NHQC) \cite{sjoqvist2012non,xu2012nonadiabatic}, which can combine advantages of both gate speed and universality, and it is currently essential as the coherent times of typical solid-state quantum systems are still limited. Therefore, after the original NHQC scheme based on three-level quantum systems \cite{sjoqvist2012non,xu2012nonadiabatic}, many alternatives of NHQC have been proposed theoretically \cite{johansson2012robustness,spiegelberg2013validity,xu2014universal, mousolou2014non, zhang2014quantum,xu2014protecting,liang2014nonadiabatic, gurkan2015realization, zhou2015cavity, xue2015universal, pyshkin2016expedited} and demonstrated experimentally \cite{abdumalikov2013experimental, feng2013experimental, zu2014experimental, arroyo2014room, danilin2018experimental}, in various quantum systems. However, earlier NHQC schemes required concatenating two separate cycles to realize arbitrary single-qubit gates, which doubled the exposure time of the quantum system to its environment, and thus increase the decoherence-induced gate error. To simplify the strategy, the single-loop/shot protocol of NHQC was proposed \cite{xu2015nonadiabatic, sjoqvist2016nonadiabatic,  zhao2017single, herterich2016single,hong2018implementing,xing2020nonadiabatic} and experimentally demonstrated in various quantum systems \cite{li2017experimental, zhou2017holonomic, sekiguchi2017optical, xu2018single, ishida2018universal, wu2022unidirectional}, which can realize arbitrary single-qubit gate with only one cyclical evolution, and thus shorten the gate time.
Besides, the NHQC has also been extended to other cases.

Due to the cyclic evolution condition in obtaining geometric phases, a holonomic quantum gate usually needs a much longer gate-time, compared with its dynamical counterpart. Therefore, it is necessary to further speed up the evolution and thus improve the fidelity of the holonomic quantum gate. So, in Section \ref{sectionFast}, we review different methods for speeding up the gate for NHQC. Firstly, incorporating  the time-optimal control technology, by solving the quantum brachistochrone equation \cite{carlini2012time, carlini2013time, wang2015quantum, geng2016experimental}, into the NHQC scheme is straightforward \cite{chen2020robust, liu2020brachistochrone, shen2021ultrafast, sun2021one}, which can optimize the  gate time under the set conditions. Secondly, as a longer evolution path usually means more gate time, various methods for shortening the evolution path are also proposed \cite{xu2018path, zhao2020general, liang2022composite, tang2022fast}. Besides, the shortcut to adiabaticity (STA) \cite{guery2019shortcuts} technique  is a successful method  for fast quantum dynamics, and thus is also proposed to be incorporated into NHQC \cite{zhang2015fast, huang2018shortcut, kang2018nonadiabatic, kang2020flexible, liu2020leakage} with experimental verifications \cite{yan2019experimental,li2022dynamical}.

It is well-known that high-fidelity and robust solutions are necessary conditions for quantum computation. As discussed in Section \ref{robustness},  benefiting from the compatibility of NHQC, the fidelity and robustness of quantum gates can be further improved by incorporating error suppression and optimization techniques. Decoherence-free subspaces (DFS) \cite{liang2014nonadiabatic, xu2014universal, zhou2015cavity, xue2015universal, pyshkin2016expedited, xue2016nonadiabatic, hu2016multi, sun2016non, zhao2016nonadiabatic, song2016shortcuts, lin2017holonomic, liu2017universal, wang2018single, ji2019scalable, chen2020robust} is the preferred solution for resisting the collective environment-induced dephasing. Other solutions, such as noiseless subsystems \cite{zhang2014quantum} and dynamical decoupling \cite{xu2018path,zhao2021dynamical}, can also provide similar results.
To enhance the robustness of systematic Rabi error, researchers have proposed optimization schemes such as pulse shaping\cite{liu2019plug, kang2019one, kang2020heralded, li2020fast,guo2020optimized, wu2021systematic, li2021multiple, liu2022optimized}, composite pulse \cite{xu2017composite,zhu2019single}, and dynamically corrected gates\cite{li2021dynamically,liu2021super,he2021robust}, etc.

To sum up, NHQC is a highly promising approach to quantum computation with the potential to enhance fault-tolerance and improve resistance against certain types of errors. Therefore, the development of NHQC has garnered great interest within the quantum computation community. Currently, we already see indications that NHQC can outperform  conventional dynamical quantum computation strategies. Despite its potential, NHQC is still in its nascent stages, and there are several challenges that must be overcome before its full potential can be realized.
\section{Adiabatic holonomic quantum computation}
\label{sectionAHQC}

\subsection{ The adiabatic non-Abelian geometric phase  }

In its early stage, HQC is a strategy  that ensures flexible information processing through all-geometric and adiabatic control. It encodes quantum information in a set of degenerate eigenstates of the Hamiltonian that depends on parameters, and then adiabatically drives these states to evolve cyclically in the  parametric space. This results in quantum computation being robust to control errors as transformations within the eigenspace are geometric and rely on the global properties of the evolution path.

We consider a Hamiltonian, denoted by $\mathcal{H}_{\lambda(t)}$, which depends on a control parameter $\lambda(t)$ and has $R$ distinct eigenvalues $\{\varepsilon_i\}_{i=1}^R$, each with a degeneracy of $\{n_i\}$. It spectral $\lambda$-dependent resolution can be written as $\mathcal{H}_{\lambda(t)}=\sum_{i=1}^R\varepsilon_i(\lambda)\prod_i(\lambda)$, where $\prod_i(\lambda)$ denotes the projector over the eigenspace spanned by $\{|\Phi_i^j(\lambda)\rangle\}_{j=1}^{n_i}$, with a common eigenvalues $\varepsilon_i(\lambda)$.
  The evolution state $|\Psi (t)\rangle$ satisfies the time-dependent Schr\"{o}dinger equation
\begin{eqnarray}
\label{Eq1}
i\frac{\partial}{\partial t}|\Psi (t)\rangle=\mathcal{H}_{\lambda(t)}|\Psi (t)\rangle.
\end{eqnarray}
When the control parameter $\lambda(t)$ is restricted to adiabatic and cyclic conditions, i.e, $\hbar\dot{\lambda}/\lambda\ll{\rm min}_{i\neq j}|\varepsilon_j-\varepsilon_i|$ and $\lambda(0)\! =\! \lambda(T)$, any initial state $|\Psi(0)$ will be mapped, after the period $T$, onto $|\Psi(T)=U(T)|\Psi(0)$, where $U(T)=\bigoplus_{l=1}^R e^{iD_l(T)}\Gamma_{A_l}(\lambda)$. Here $D_l(T)=\int_0^T\varepsilon_l(\lambda_t)dt$ is the dynamical phase, and
\begin{eqnarray}
\label{Eq2}
\Gamma_{A_l}(\lambda)=P{\rm exp}\int_\lambda A_l\in U(n_l) \  \ \quad (l=1,...,R)
\end{eqnarray}
is the holonomy, with $P$ being the path ordering. The elements of $A_l$ are
\begin{eqnarray}
\label{Eq3}
(A_{l,\mu})^{\alpha,\beta}=\langle \Phi_l^{\alpha}(\lambda)|\frac{\partial}{\partial\lambda^{\mu}}|\Phi_l^{\beta}(\lambda)\rangle,
\end{eqnarray}
where $(\lambda_{\mu})_{\mu=1}^d$ are the local coordinates on the control parameter space. For $n_l>1$ the holonomy $\Gamma_{A_l}(\lambda)$ is referred to non-Abelian geometric phase \cite{wilczek1984appearance,zanardi1999holonomic}.

\subsection{Holonomic quantum gates}\label{adiabatic HQC}

Adiabatic HQC was proposed in trapped ions systems \cite{duan2001geometric, kuvshinov2005robust, kuvshinov2006decoherence}, which controlled transitions between four energy levels by using a tripod configuration with three separate laser pulses. Later, it had been expanded to various systems, including atomic/atomic ensemble \cite{recati2002holonomic, li2004non, moller2007geometric, zheng2012geometric}, superconducting \cite{siewert2002holonomic, faoro2003non, cholascinski2004quantum, zhang2005holonomic, feng2008holonomic, brosco2008non, pirkkalainen2010non, kamleitner2011geometric, chancellor2013scalable}, Bose-Einstein condensates \cite{fuentes2002geometric}, semiconductor  systems \cite{solinas2003holonomic, solinas2003semiconductor, bernevig2005holonomic, parodi2006fidelity, golovach2010holonomic, budich2012all}, spin chains \cite{karimipour2005exact, ota2008implementation, renes2013holonomic}, neutral particle \cite{bakke2011quantum} and optical system \cite{pinske2020highly}. In addition to the four-level tripod model, other models can also achieve adiabatic HQC \cite{heydari2012combinatorial, malinovsky2014adiabatic}. To improve its fault tolerance performance, one can encode the information into subsystems \cite{nordling2005mixed, oreshkov2009holonomic}, adopt the error correction coding\cite{wu2005holonomic, zhang2006physical, oreshkov2009fault, oreshkov2009scheme, zheng2014fault, albert2016holonomic}, and expand adiabatic HQC into the topological regime \cite{calzona2020holonomic, zhang2020topological}.
Geometric phases in open quantum systems have been examined when considering the interaction between a quantum system and its environment \cite{fuentes2005holonomic, pinske2019holonomic}, and the  robustness has been analyzed \cite{kuvshinov2003stability, solinas2004robustness, florio2006robust, trullo2006robustness}.

Although  the theory of adiabatic HQC has been well developed, its implementation  remains challenging due to difficulties in controlling the complex energy level structure and achieving high-quality manipulation under the required slow adiabatic evolution. So far, there have been very few experiments that have achieved adiabatic holonomic gates\cite{toyoda2013realization, leroux2018non}. In Ref. \cite{toyoda2013realization}, single-qubit adiabatic holonomic gates can be achieved by using a four-level system of trapped $^{40}{\rm Ca^+}$ ion that was connected by three oscillating fields.
Using quantum state tomography, the fidelities of $x$ and $z$ gates with $\pi$ rotation, and the Hadamard gate were estimated to be $0.965$, $0.931$, and $0.965$, respectively. The infidelities of quantum gates were largely due to imperfect initialization and analysis. The study also explored its robustness against variations in parameters.
In addition, Ref. \cite{leroux2018non} realized non-Abelian ${\rm SU(2)}$ geometrical transformations acting on the dark states of the system and demonstrated the non-Abelian characteristics by cycling the relative phase of the tripod beams in the strontium-87 atom laser cooling gas.

\section{Nonadiabatic holonomic quantum computation}\label{NHQC}

To relieve the constraint imposed by the adiabatic condition, which slows down the speed of evolution, the NHQC was proposed \cite{sjoqvist2012non, xu2012nonadiabatic}
using a three-level quantum system, and has since become the subject of numerous studies and evaluations.

\subsection{ Nonadiabatic non-Abelian geometric phase }

We consider an $N$-dimensional quantum system controlled by the Hamiltonian $H(t)$. Suppose there is a time-dependent $L$-dimensional subspace $S(t)={\rm Span}\{|\varphi_k(t)\rangle\}_{k=1}^L$ undergoes the cyclic evolution around a smooth path $C$, i.e., $S(\tau)=S(0)$. Here $|\varphi_k(t)\rangle$ satisfy the Schr\"{o}dinger equation $i|\dot{\varphi}_k(t)\rangle=H(t)|\varphi_k(t)\rangle$. We can introduce a set of auxiliary vectors $\{|\nu_k(t)\rangle\}_{k=1}^L$ of $S(t)$ along the smooth path $C$, with the property that $|\nu_k(\tau)\rangle=|\nu_k(0)\rangle=|\varphi_k(0)\rangle$, which do not have to be solutions of the Schr\"{o}dinger equation. Therefore, the evolution states $|\varphi_k(t)\rangle$ can be expressed as
\begin{eqnarray}
\label{Eq14}
|\varphi_k(t)\rangle=\sum_{l=1}^L|\nu_l(t)\rangle \mathcal{C}_{lk}(t), \ k=1,2,...,L,
\end{eqnarray}
where $\mathcal{C}_{lk}(t)$ are the time-dependent coefficients. The unitary evolution operator at $\tau$ becomes
\begin{eqnarray}
\label{Eq15}
U(\tau)= \sum_k|\varphi_k(\tau)\rangle\langle\varphi_k(0)| =\sum_{l,k=1}^L\mathcal{C}_{lk}(\tau)|\nu_l(0)\rangle\langle\nu_k(0)|,
\end{eqnarray}
which means that at the moment when the condition of cyclic evolution is met, the matrix element of the evolution operator is $\mathcal{C}_{lk}(\tau)$.
Moreover, combining Eq.~(\ref{Eq14}) and the Schr\"{o}dinger equation, we can obtain
\begin{eqnarray}
\label{Eq16}
\dot{\mathcal{C}}_{lk}(t)=i\sum_{m=1}^L[\mathcal{A}_{lm}(t)-\mathcal{K}_{lm}(t)] \mathcal{C}_{mk}(t).
\end{eqnarray}
Then,
\begin{eqnarray}
\label{Eq17}
U(\tau)=\mathcal{C}(\tau)=T{\rm exp}\left\{{i\int_0^{\tau}[\mathcal{A}(t)-\mathcal{K}(t)]dt}\right\},
\end{eqnarray}
with $\mathcal{A}_{lm}(t)=i\langle\nu_l(t)|\dot{\nu}_m(t)\rangle$ and $\mathcal{K}_{lm}(t)=\langle\nu_l(t)|H(t)|\nu_m(t)\rangle$ being elements of $L \times L$  Hermitian matrices. Here $\mathcal{A}_{lm}(t)$ represents the geometric component because it is independent of the Hamiltonian and only depends on the Hilbert space structure, and $\mathcal{K}_{lm}(t)$ denotes the dynamical component.

To understand the meaning of $\mathcal{A}$, we choose another set of auxiliary vectors $\{|\nu'_k(t)\rangle\}_{k=1}^L$, with $|\nu'_k(\tau)\rangle=|\nu'_k(0)\rangle=|\varphi_k(0)\rangle$. There is an unitary operator $\mathcal{V}(t)$, which is constrained by the boundary condition $\mathcal{V}(\tau)=\mathcal{V}(0)=I$, can make the transformation of $|\nu'_k(t)\rangle=\sum_{l=1}^L|\nu_l(t)\rangle \mathcal{V}_{lk}(t)$. By using these new vectors, we can obtain $\mathcal{A}'_{lm}(t)=i\langle\nu'_l(t)|\dot{\nu}'_m(t)\rangle$, which meets $\mathcal{A}'_{lm}=(\mathcal{V}^{\dag}\mathcal{A}\mathcal{V})_{lm} +i(\mathcal{V}^{\dag}\dot{\mathcal{V}})_{lm}$. This indicates that $\mathcal{A}$ transforms as a proper vector potential, and $\mathcal{A}_{lm}$ is the element of holonomy matrix  generalizing the Wilczek¨CZee holonomy \cite{wilczek1984appearance} to the nonadiabatic case.

\subsection{The NHQC scheme}

To construct universal gates with speed and built-in fault-tolerant features, the NHQC based on the nonadiabatic non-Abelian geometric phases was proposed. For universal NHQC, the following cyclic evolution condition and parallel transport condition are necessary \cite{xu2012nonadiabatic,sjoqvist2012non}, i.e.,
\begin{eqnarray}
\label{Eq12}
\sum_{k=1}^L|\varphi_k(\tau)\rangle\langle\varphi_k(\tau)| =\sum_{k=1}^L|\varphi_k(0)\rangle\langle\varphi_k(0)|, \quad
\langle\varphi_k(t)|H(t)|\varphi_l(t)\rangle=0,\ k,l=1,2,...,L,
\end{eqnarray}
In these conditions, the unitary transformation $U(\tau)$, as defined in equation (\ref{Eq17}), takes the form $U(\tau)=T{\rm exp}[{i\int_0^{\tau}\mathcal{A}(t)dt}]$. This transformation is a nonadiabatic holonomic gate that acts on the $L$-dimensional computational subspace $S(0)$. It is clear that the evolution operator $U(\tau)$ acting on the computational space depends only on the subspace determined by $\{|\varphi_k(t)\rangle\}_{k=1}^L$, and not on the selection of auxiliary vectors. From another perspective, $U(\tau)$ depends solely on the evolution path of the quantum state, and not on the specific details of the evolution, which makes it resistant to various types of control errors. This property enables us to construct universal gates with both speed and built-in fault-tolerant features.

\begin{figure}[t]
  \centering
  \includegraphics[width=0.7\linewidth]{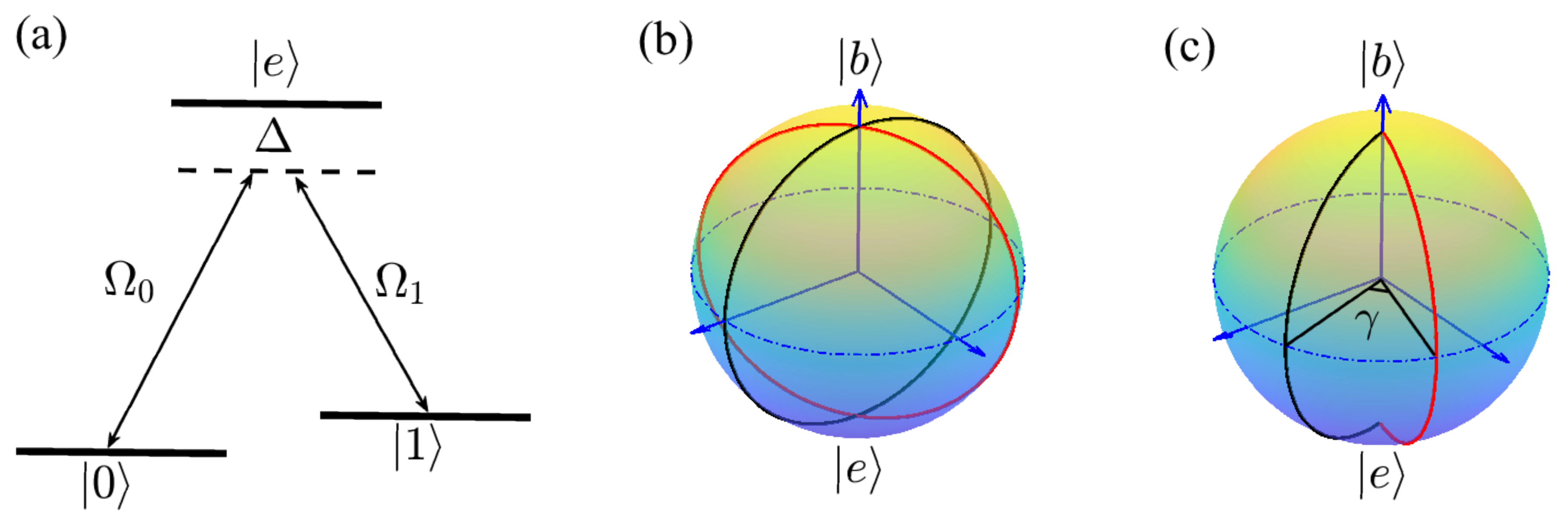}
  \caption{(a) Illustration of the construction of single-qubit nonadiabatic holonomic quantum gates, with $|0\rangle$ and $|1\rangle$ representing the encoding space, and state $|e\rangle$ representing the auxiliary level. The detuning, represented by $\Delta$, is zero in the case of resonant coupling. (b) Evolution path with two loops. (c) Evolution path with a single loop, with $\gamma$ representing the geometric phase.}
  \label{Fig4}
\end{figure}

The initial NHQC schemes \cite{sjoqvist2012non, xu2012nonadiabatic} implemented by using the resonance three-level model, as illustrated in Fig.~\ref{Fig4}(a), with $\Delta=0$. The qubit consists of two ground states, denoted as $|0\rangle$ and $|1\rangle$, with an excited state labeled as $|e\rangle$. The corresponding driving fields are denoted as $\Omega_0(t)=\Omega_r(t)\omega_0$ and $\Omega_1(t)=\Omega_r(t)\omega_1$, respectively. In the interaction picture, the Hamiltonian is expressed as
\begin{eqnarray}
\label{resonance}
H_{r}(t)&=&\Omega_0(t)|e\rangle\langle 0|+\Omega_1(t)|e\rangle\langle 1|+\rm{H.c.}\notag \\
&=&\Omega_r(t)|e\rangle\langle b|+\rm{H.c.},
\end{eqnarray}
where $|b\rangle=\omega_0^*|0\rangle+\omega_1^*|1\rangle$ is the bright state.
Then, the evolution of the quantum system can be viewed as a Rabi oscillation between the bright state $|b\rangle$ and the excited state $|e\rangle$, with a Rabi frequency of $\Omega_r(t)$. The dark state $|d\rangle=-\omega_1 |0\rangle+\omega_0|1\rangle$ is decoupled from the dynamics. When the cyclic evolution condition $\int_0^{\tau}\Omega_r(t')dt'=\pi$ is satisfied, the system completes a cyclic evolution and returns to the computation space spanned by $|0\rangle $ and $|1\rangle $. By setting $\omega_0=\sin(\theta/2)e^{i\phi}$ and $\omega_1=-\cos(\theta/2)$, the evolution operator at the final moment is expressed as $ U(\tau)=\exp (-i \pi \textbf{ n}\bm{\cdot\sigma}/2)$, which is a rotation operation around the axis $\textbf{ n}=(\sin\theta \cos\phi,\ \sin\theta \sin\phi, \ \cos\theta)$ by an angle $\pi$. However, an arbitrary single-qubit gate needs to be constructed by two sequential rotations here, as the obtained geometric phase is fixed. As depicted in Fig.~\ref{Fig4}(b), the starting and ending points of each loop are the bright state $|b\rangle$.

The nonadiabatic holonomic quantum computation has shown great promise as a method for quantum computation in reducing excessively long evolution times and withstanding local fluctuations. This has been experimentally verified in a superconducting circuit \cite{abdumalikov2013experimental} and a liquid nuclear magnetic resonance (NMR)  system \cite{feng2013experimental}, where single-qubit gates have been performed with fidelities exceeding $95\%$. They also provided evidence for the non-Abelian character of the implemented holonomic quantum operations. Besides, the holonomic  controlled NOT gates with an average fidelity of  $93.12\%$ is also demonstrated in Ref. \cite{feng2013experimental}.
The experimental demonstrations of NHQC gates in superconducting qubit and liquid NMR systems are significant milestones in the development of NHQC. Afterwards, NHQC has been successfully implemented in various physical systems \cite{zu2014experimental, arroyo2014room, danilin2018experimental,xu2021demonstration,egger2019entanglement}.

\subsection{The single-shot/loop NHQC}
In early NHQC schemes, two separate continuous evolution operators were necessary to perform an arbitrary single-qubit gate, leading to increased errors due to decoherence. To remove this obstacle, researchers have developed improved methods that allow for the realization of arbitrary single-qubit gates through a single-loop evolution.

One such approach is to utilize detuned laser pulses in a three-level $\Lambda$ system to achieve quantum gates, known as single-shot NHQC \cite{xu2015nonadiabatic, sjoqvist2016nonadiabatic, zhao2017single, xing2020nonadiabatic}. As illustrated in Fig.~\ref{Fig4}(a), the transitions between ground states and the excited state are driven by $\Omega_0(t)$ and $\Omega_1(t)$, with a detuning of $\Delta$. In the interaction picture, the Hamiltonian can be written as
\begin{eqnarray}
\label{single shot}
H_{ss}(t)&=&[\Omega_0(t)|0\rangle\langle e|+\Omega_1(t)|1\rangle\langle e|+\rm{H.c.}]
-\Delta|e\rangle\langle e|.
\end{eqnarray}
By setting $\Omega_0(t)=\Omega_{ss}\cos\alpha\cos\gamma$, $\Omega_1(t)=\Omega_{ss} e^{i\beta}\sin\alpha\cos\gamma$, and $\Delta=-2\Omega_{ss}\sin\gamma$,
the Hamiltonian becomes
\begin{eqnarray}
\label{single shot1}
H_{ss}(t)=\Omega_{ss}\sin\gamma(|e\rangle\langle e|+|b\rangle\langle b|)+\Omega_{ss}[\cos\gamma(|e\rangle\langle b|+|b\rangle\langle e|) +\sin\gamma(|e\rangle\langle e|-|b\rangle\langle b|)],
\end{eqnarray}
where $|b\rangle=\cos\alpha|0\rangle+e^{i\beta}\sin\alpha|1\rangle$ is the bright state, and $|d\rangle=\sin\alpha|0\rangle-e^{i\beta}\cos\alpha|1\rangle$ is the dark state decoupled from the dynamics.  Here $\alpha$, $\beta$, and $\gamma$ are time-independent parameters. When the evolution period $T=\pi/\Omega_{ss}$, the evolution operator under the basis vector $|b\rangle$, $|e\rangle$  and $|d\rangle$ can be written as $U_{ss}(T)=e^{-i\phi}|e\rangle\langle e|+e^{-i\phi}|b\rangle\langle b|+|d\rangle\langle d|$, with $\phi=\pi\sin\gamma+\pi$. In the computation space spanned by $|0\rangle$ and $|1\rangle$, the evolution operator can be recast as $U(T)=\exp[-i \phi (|b\rangle\langle b|-|d\rangle\langle d|)/2]$.

It is important to note that both $\Omega_{0,1}(t)$ and $\Delta$ are dependent on the frequency and amplitude of the laser. This results in a time-dependent synchronous variation between the laser fields and the detuning, making it difficult for pulse waveforms other than square waves to meet these requirements. This is one  limitation of this scheme.  Furthermore, due to the restriction $\tan\gamma=\Delta/(-2\sqrt{|\Omega_0|^2+|\Omega_1|^2 })$, the rotation angle in the evolution operator can only be within a specific range limited by the system parameters, which is the other limitation of the scheme.

The single-shot NHQC was first demonstrated  experimentally  in a NMR system \cite{li2017experimental}. They constructed the $R_x (\pi/2)$, $R_z (\pi/2)$, $R_x (\pi)$ and $R_z (\pi)$ gates and tested the corresponding gate fidelities, which were $98.07\%, 98.29\%, 99.68\%$ and $99.75\%$, respectively, using quantum state tomography (QST). Subsequently, arbitrary single-qubit gates were obtained in a nitrogen-vacancy center in diamond \cite{zhou2017holonomic}. However, the gate fidelities there are limited by additional errors, such as excited state occupation and crosstalk between driving fields. Additionally, all single-qubit Clifford gates were experimentally demonstrated in a three-level superconducting Xmon qutrit \cite{zhang2019single}. Characterized by both QST and randomized benchmarking (RB), all gate fidelities exceed $99\%$. Furthermore, the geometric spin in a degenerate subspace of a spin-1 electronic system under zero field in a nitrogen-vacancy center in diamond can also be used as a platform for implementing the single-shot NHQC \cite{sekiguchi2017optical}.

The single-shot method may simplify the original two-loop NHQC scheme somewhat, but it is not compatible with pulse shaping and presents some practical difficulties. Hence, the single-loop NHQC (SL-NHQC) \cite{herterich2016single, hong2018implementing} is a better alternative, as it is compatible with various optimization techniques. This is made possible by using resonant-driven multiple pulses. The key idea behind the multi-pulse method is to divide a single-loop into segments to generate a holonomic gate. The Hamiltonian in this multi-pulse approach is
\begin{eqnarray}
\label{sl}
H_{sl}(t)&=&\Omega_{0}(t)e^{-i\phi_0}|0\rangle\langle e|+\Omega_{1}(t)e^{-i\phi_1}|1\rangle\langle e|+\rm{H.c.}\notag \\
&=&\Omega_{sl}(t)e^{-i\phi_0}|b\rangle\langle e|+\rm{H.c.},
\end{eqnarray}
with $\Omega_{sl}(t)= \sqrt{\Omega_{0}(t)^2+\Omega_{1}(t)^2}$. The dynamics of the system is
equivalent to a resonant coupling between the bright state $|b\rangle=\sin( \theta/2)|0\rangle-\cos( \theta/2)e^{i\phi}|1\rangle$ and the excited state $|e\rangle$, while the dark state $|d\rangle=\cos( \theta/2)e^{-i\phi}|0\rangle+\sin( \theta/2)|1\rangle$ is decoupled from the dynamics of the system, with $\tan( \theta/2)=\Omega_{0}(t)/\Omega_{1}(t)$, and $\phi=\phi_0-\phi_1+\pi$.

We divide the evolution period into two equal intervals \cite{herterich2016single,hong2018implementing}. In the first interval $t\in[0,T/2]$, we set $\phi_0=0$, which reduces the Hamiltonian to $H_1=\Omega_{sl}(t)(|b\rangle\langle e|+|e\rangle\langle b|)$. Under the condition that $\int^{T/2}_0\Omega_{sl}(t)dt=\pi/2$, the evolution operator in the first interval is $U_1=|d\rangle\langle d|-i(|b\rangle\langle e|+|e\rangle\langle b|)$. In the second interval $t\in[T/2,T]$, the parameter turns to be $\phi_0=\pi-\gamma$, and the corresponding Hamiltonian is $H_2=-\Omega_{sl}(t)(e^{i\gamma}|b\rangle\langle e|+e^{-i\gamma}|e\rangle\langle b|)$. With the condition $\int^{T}_{T/2}\Omega_{sl}(t)dt=\pi/2$, the evolution operator is given by $ U_2=|d\rangle\langle d|+i(e^{i\gamma}|b\rangle\langle e|+e^{-i\gamma}|e\rangle\langle b|)$. At the final moment, the evolution operator in the computation space is obtained as
\begin{eqnarray}
\label{slu}
U_{sl}(T)=\left(
\begin{array}{cccc}
 \cos\frac {\gamma} {2}-i\sin\frac {\gamma} {2}\cos\theta          & -i\sin\frac {\gamma} {2} \sin\theta e^{i\phi}\\
 -i\sin\frac {\gamma} {2} \sin\theta e^{-i\phi} & \cos\frac {\gamma} {2}+i\sin\frac {\gamma} {2}\cos\theta
\end{array}
\right)
={\rm exp}(-i\frac{\gamma}{2}\textbf{n}\cdot\bm{\sigma}),
\end{eqnarray}
which is a rotation operation around the axis $\textbf{n}=(\sin\theta \cos\phi,\ \sin\theta \sin\phi, \ \cos\theta)$ by an angle of $\gamma$. Thus, the arbitrary single-qubit gates can be constructed by choosing different values of parameters $\theta, \phi$, and $\gamma$. The corresponding evolution path is just a single-loop on the Bloch sphere, as depicted in Fig.~\ref{Fig4}(c).

\begin{figure}[!t]
  \centering
  \includegraphics[width=0.6\linewidth]{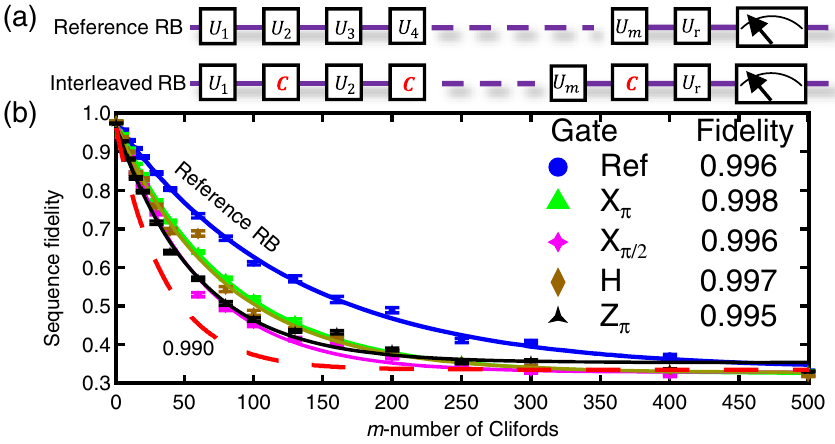}
  \caption{Randomized benchmarking (RB) of the single-qubit holonomic gates on a transmon qubit. (a) Sequence of both a reference RB experiment and an interleaved RB experiment. (b) Sequence fidelity as a function of the gate length m. Adapted with permission \cite{xu2018single}. Copyright 2018, American Physical Society.} \label{Fig6}
\end{figure}

Single-qubit gates in the single-loop scheme have been experimentally demonstrated  in both superconducting transmon qubit and a microwave photonic qubit \cite{xu2018single}. As shown in Fig.~\ref{Fig6}, characterized by RB, the average gate fidelity in the transmon qubit is found to be 99.6$\%$, which represents a significant improvement compared to single-shot NHQC. In addition to single-qubit gates, nontrivial two-qubit gates of NHQC have also been demonstrated in geometric spin qubits \cite{nagata2018universal}.

\subsection{Extensions of NHQC} \label{extension}

The conventional NHQC must adhere to two conditions: the cyclic evolution condition and the parallel transport condition. This results in a decrease in the freedom of parameters, ultimately limiting its widespread practical applications.
For maintaining the geometric robustness of NHQC and relaxing the restrictions on parameters,  researchers have made some extensions.

Firstly, the extension to the unconventional NHQC case, which is based on the unconventional geometric phase, eases the restriction of the parallel transport condition. This is because it allows for a non-zero dynamical phase by requiring the dynamical phase to be proportional to the geometric phase, thereby maintaining the same  geometric properties as the pure geometric phase. The unconventional geometric phase was first introduced  \cite{zhu2003unconventional} to create a geometric gate, and has then been expanded to include the implementation of NHQC \cite{chen2020robust, shen2021ultrafast, liu2020brachistochrone, ji2022nonadiabatic}.
From the evolution operator in Eq.~(\ref{Eq17}), we can choose another set of auxiliary vectors $\{|\nu_k[\lambda_a(t),\eta_b(t)]\rangle\}_{k=1}^L$, which has two sets of independent parameters $\lambda_a(t)$ and $\eta_b(t) \ (a,b=1,2,...,n)$. Then, the geometric component $\mathcal{A}_{lm}(t)$ becomes $\mathcal{A}_{lm}(t)=\mathcal{A}_{lm}^{\lambda}+\mathcal{A}_{lm}^{\eta}$, where $\mathcal{A}_{lm}^{\lambda}=\sum_a i\langle\nu_l(t)|\frac{\partial}{\partial \lambda_a}| \nu_m(t)\rangle(d\lambda_a/dt)$ and $\mathcal{A}_{lm}^{\eta}=\sum_b i\langle\nu_l(t)|\frac{\partial}{\partial \eta_b}| \nu_m(t)\rangle(d\eta_b/dt)$. We set  the dynamical part $K_{lm}(t)$ to be offset by the geometric part $A_{lm}^{\eta}$ at each moment, i.e., $K_{lm}+\mathcal{A}_{lm}^{\eta}=0$. As a result, the evolution operator in Eq.~(\ref{Eq17}) becomes $U(\tau)=\mathcal{P}{\rm exp}\left (i\oint A^{\lambda_a}d\lambda_a \right)$, which is an unconventional holonomy. Here $\mathcal{P}$ is the path ordering along the closed path, and $\mathcal{A}^{\lambda_a}$ is the non-Abelian connection. In this case, the dynamical component  $K_{lm}(t)\neq 0$, so we call it unconventional nonadiabatic holonomy.

On the other hand, due to the limitation of the cyclic evolution, the NHQC has a strict limit on the evolution time. Therefore, researchers extended it to the non-cyclic evolution case to relax the condition of cyclic evolution \cite{li2021noncyclic}. With the help of dynamical invariants, they obtained the evolution operator at the last moment as $U_{nc}(T)=\sum_{n=0,\pm}e^{i\alpha_n}|\varphi_n(T)\rangle\langle\varphi_n(0)|$, where $|\varphi_n(t)\rangle $ are the eigenstates of invariant, and $\alpha_n$ are the Lewis-Riesenfeld phase. The key idea of this approach is to divide the evolution path into two equal parts and change the driving parameters at the midpoint to swap the two orthotropic channels $|\varphi_{\pm}(t)\rangle$ so as to remove the dynamical phase.

Note that, although the three-level $\Lambda$ system is the typical building block for NHQC, exploring NHQC beyond the three-level setting can expand the Hilbert space and reduce the circuit complexity  for quantum algorithm \cite{xu2021realizing, ai2022experimental, andre2022dark}.

\section{Speeding up holonomic quantum gates}
\label{sectionFast}

High-fidelity quantum gates are a necessary component for achieving quantum computation. However, quantum systems are inevitably impacted by environment-induced decoherence, causing a reduction in gate fidelity. While NHQC can improve the gate fidelity by using pulses that are short compared to the decay time scale, this requires extremely high pulse intensities and rapid pulse changes, making it challenging to implement experimentally. Hence, to improve the fidelity of the holonomic gate within the range of pulse intensities that can be achieved experimentally, new optimization methods have been explored, such as time-optimal control NHQC and short-path NHQC.

\subsection{NHQC with time-optimal control technique}

Although the NHQC scheme reduces the evolution time to some extent, it is still important to further shorten it to minimize the decoherence-induced errors because the coherence times of quantum systems are limited. By solving the quantum brachistochrone equation \cite{carlini2012time, carlini2013time, wang2015quantum, geng2016experimental}, the time-optimized technology provides a way to achieve the NHQC with the shortest time \cite{chen2020robust, shen2021ultrafast, liu2020brachistochrone}. However, due to additional constraints, the resulting geometric phase is of an unconventional nature as described in Section ~\ref{extension}. Thus, we call this scheme as time-optimal unconventional NHQC (TO-UNHQC). The shortest time can be obtained by solving the time-dependent Schr\"{o}dinger equation in conjunction with the quantum brachistochrone equation, i.e.,
 \begin{eqnarray}
\label{Eq18}
i\partial F/\partial t=[H_T,F],
\end{eqnarray}
where $F=\partial L_c/\partial H_T$, $L_c=\sum_j\mu_jf_j(H_T)$, and $\mu_j$ is the Lagrange multiplier.

\begin{figure}[t]
  \centering
  \includegraphics[width=0.5\linewidth]{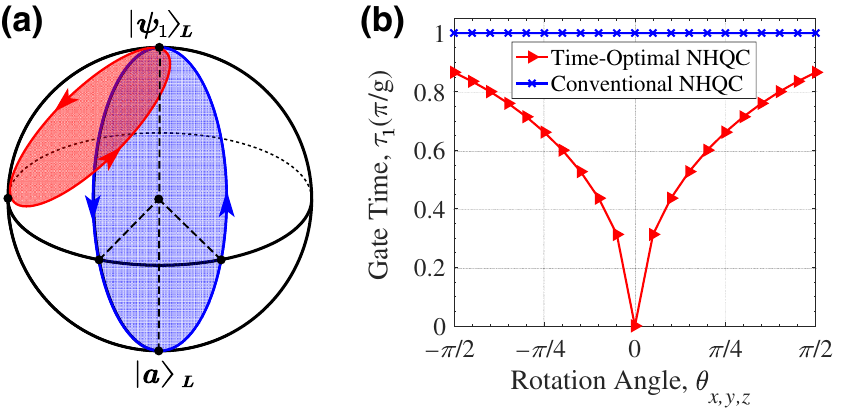}
  \caption{(a) The evolution paths of TO-UNHQC (red line) and conventional NHQC (blue line) schemes. (b) Comparison of time costs as a function of rotation angles. Adapted with permission \cite{chen2020robust}. Copyright 2020, American Physical Society.} \label{TOC}
\end{figure}

To implement the TO-UNHQC, we consider a three-level system as shown in Fig.~\ref{Fig4}(a). The Hamiltonian of the system is
\begin{eqnarray}
\label{HTO}
H_{T}(t)=\frac{\Omega_T(t)}{2}e^{-i\phi(t)}(\sin\frac{\theta}{2}e^{i\phi_1}|0\rangle +\cos\frac{\theta}{2}|1\rangle)\langle e|+\rm H.c.
\end{eqnarray}
In the physical system, the energy bandwidth must also be taken into account, meaning that the condition $f_1(H_T)\equiv\frac{1}{2}[Tr(H_T^2)-\frac{\Omega(t)^2}{2}]=0$ needs to be satisfied.
Additionally, since the independent $\sigma_z$ operation cannot be implemented directly in the experiment, the constraint of $f_2(H_T)\equiv Tr(H_T\sigma_z)=0$ must also be satisfied.
By solving the Schr\"{o}dinger equation and quantum brachistochrone equation, and setting $\Omega_T(t)=\Omega_0$, we can obtain the minimum-time solution to this method as  $\phi(t)=2(\gamma-\pi)t/\tau$, with a minimum evolution time $\tau=2\sqrt{\pi^2-(\pi-\gamma)^2}/\Omega_0$. This evolution time decreases as the geometric phase $\gamma$ decreases. As shown in Fig.~\ref{TOC}, the TO-UNHQC scheme possesses the shortest evolution time and shorter evolution path compared to the conventional NHQC.

Recently, the TO-UNHQC has been successfully implemented in an Xmon-type superconducting circuit  \cite{han2020experimental}. The experiment results demonstrate the superiority of TO-UNHQC in terms of gate operation time and its lower sensitivity to detuning error compared to conventional NHQC. They achieved a fidelity of $99.51\%$ for a single qubit gate using interleaved RB. Moreover, they implemented a two-qubit holonomic control-phase gate, and proved it more robust against certain control noises than the conventional SL-NHQC.
Subsequently, fast and high-fidelity implementation of NHQC using the time-optimal method was proposed in a hybrid spin register in diamond  \cite{dong2021fast}. With the help of time-optimal control, the gate-fidelities of the single-qubit gate and two-qubit gate there can exceed $99.2\%$ and $96.5\%$, respectively.

\subsection{NHQC with path-optimization }

Path optimization is an alternative efficient way to speed up holonomic quantum gates by shortening the evolution path. A first step  for optimizing the path is made by the single-loop scheme \cite{hong2018implementing}, which reduces the two loops required to construct an arbitrary holonomic gate in the original plan to just one loop, cutting the evolution path and time in half. However, in the single-loop scheme, regardless of the type of rotation gate, the evolution path on the Bloch sphere always has to go from the North Pole to the South Pole and then back to the North Pole, resulting in the same evolution time for all rotation operations. This makes the path to be much longer than that of the corresponding dynamical gate.

To break this limitation, a shortened path scheme is proposed \cite{xu2018path}, which further shortens the evolution path of a quantum system by segmentally selecting the parameters of the non-resonant three-level Hamiltonian. The Hamiltonian in the $j$th segment is given by $H_j=\Delta_j|e\rangle\langle e|+\Omega_j[\rm {exp} (i\phi_j)|e\rangle\langle b|+\rm H.c.]$, where all the parameters are time-independent. The key to the design of the path-shortening scheme is to establish the connection conditions that must be met between the segment parameters while satisfying the constraints of the cyclic condition and the parallel evolution condition. For example, we set $\vartheta_j=\sqrt{(\Delta_j/2)^2+\Omega_j^2}\tau_j \ (j=1, 2)$, and $\tan\eta_j=2\Omega_j/\Delta_j$, with $\tau_j$ being the evolution period of the $j$th section. If the condition $|\sin\vartheta_1\sin\eta_1|=|\sin\vartheta_2\sin\eta_2|$ is met, the phase parameter $\phi_2$ can be adjusted so that the quantum system satisfies both the cyclic and the parallel evolution condition throughout the entire evolution process. The improvement of the path-shortening scheme is that it relaxes the restriction of $\Sigma_j\vartheta_j=\pi$ in the previous scheme to $\Sigma_j\vartheta_j<\pi$, thereby shortening the evolution path of the system and reducing the operation time of the quantum gate.

Subsequently, a general way to construct  NHQC is proposed \cite{zhao2020general}  via inverse engineering. This method can be used to construct the Hamiltonian that implements arbitrary $L$-dimensional holonomic quantum gates.
To implement an $L$-dimensional holonomic gate, we select a set of auxiliary vectors $\{|\mu_k(t)\rangle\}_{k=1}^{L+1}$ that satisfy the cyclic condition $|\mu_k(0)\rangle=|\mu_k(\tau)\rangle$, and set the Hamiltonian of the system to be
\begin{eqnarray}
\label{HS}
H_{S}(t)&=&\left [i\sum_{i=1}^{L}\langle \mu_{i}(t)| \dot{\mu}_{L+1}(t)\rangle |\mu _{i}(t)\rangle\langle \mu_{L+1}(t)|+\rm{H.c.}\right ]\notag\\
&& +\left [i \langle \mu_{L+1}(t)| \dot{\mu}_{L+1}(t)\rangle-\dot{\zeta}(t)\right ]|\mu _{L+1}(t)\rangle\langle \mu_{L+1}(t)|,
\end{eqnarray}
where $\zeta (t)$ is a time-dependent real function with $\zeta (0)=0$. In this case, the subspace $S_L(t)={\rm Span}\{|\phi_k(t)\rangle\}_{k=1}^L$ spanned by solutions of the Schr\"{o}dinger equation can readily fulfill the cyclic and parallel transport conditions, where $|\phi_k(t)\rangle$ are solutions of the Schr\"{o}dinger equation with initial condition $|\phi_k(0)\rangle=|\mu_{k}(0)\rangle$.
By denoting the computational subspace as $S_L(0)={\rm Span}\{|\phi_k(0)\rangle\}_{k=1}^L$, the evolution operator $U(\tau)=\mathcal{T} {\rm exp}[{i\int_{0}^{\tau}A(t)dt}]$ acting on the computational subspace is a path-dependent holonomic gate, with $A_{ij}(t)=i\langle\mu_{i}(t)|\dot{\mu}_{j}(t)\rangle$. The structure of the auxiliary vectors $\{|\mu_k(t)\rangle\}_{k=1}^{L+1}$ dictates the path of evolution in the parameter space. This implies that there are multiple paths for the same rotation operation. It is noteworthy that these paths need not necessarily traverse the South Pole, which distinguishes it from the single-loop scheme and is also crucial for path optimization.

\begin{figure}[!t]
  \centering
  \includegraphics[width=0.5\linewidth]{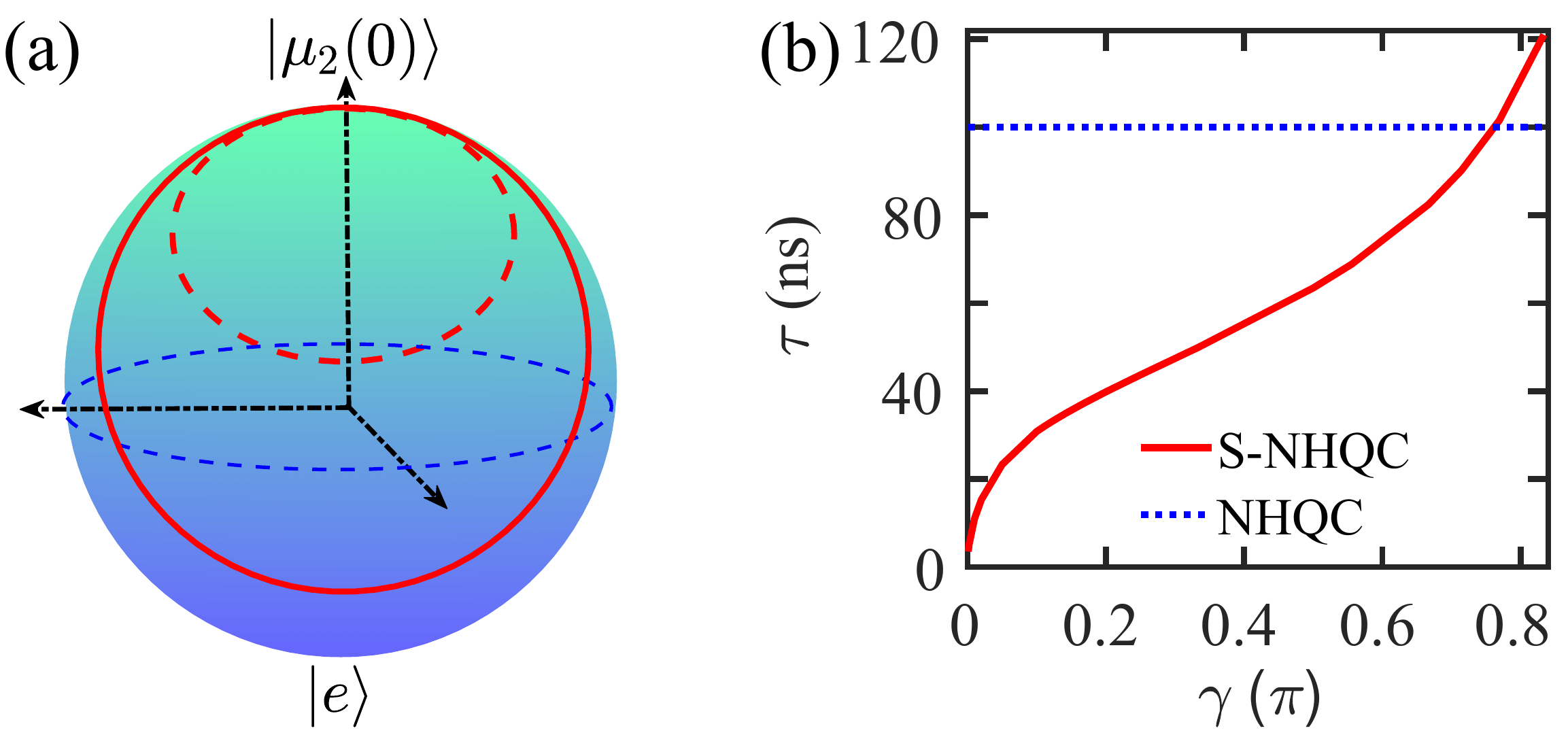}
  \caption{ (a) Evolution paths of different gates for the S-NHQC scheme, where the dashed line is the evolution path of the $T$ gate, and the solid line is the evolution path of the $S$ gate and the $\sqrt{H}$ gate. (b) The comparison of the $z$-axis-rotation gate time-costs as a function of rotation angles. Reproduced with permission \cite{liang2022composite}. Copyright 2022, American Physical Society.} \label{shortest-path}
\end{figure}

We recently constructed the shortest-path NHQC (S-NHQC) scheme \cite{liang2022composite}, as shown in Fig. \ref{shortest-path}, where the  path is a circle in the parameter space. To create arbitrary single-qubit gates, we choose a set of auxiliary vector as
\begin{eqnarray}
\label{}
|\mu_{1}(t)\rangle&=&\cos\frac{\theta}{2}|0\rangle+\sin\frac{\theta}{2}e^{i\varphi}|1\rangle, \notag\\ |\mu_{2}(t)\rangle&=&\cos\frac{\alpha(t)}{2}\left(\sin\frac{\theta}{2}e^{-i\varphi}|0\rangle -\cos\frac{\theta}{2}|1\rangle\right )+\sin\frac{\alpha(t)}{2} e^{i\beta(t)} |e\rangle, \notag\\
|\mu_{3}(t)\rangle&=&\sin\frac{\alpha(t)}{2} e^{-i\beta(t)}\left(\sin\frac{\theta}{2}e^{-i\varphi}|0\rangle-\cos\frac{\theta}{2}|1\rangle\right) -\cos\frac{\alpha(t)}{2}|e\rangle,
\end{eqnarray}
where $\theta, \varphi$ are time-independent parameters used to define the rotation axis. $\alpha(t), \beta(t)$
denote the time-dependent polar angle and azimuthal angle of
a spherical coordinate system, with $\alpha(\tau)=\alpha(0) = 0$. The subspace $S_L(0)={\rm Span}\{|\mu_{1}(0)\rangle, |\mu_{2}(0)\rangle\}$ denotes the computational space, and $|\mu_{3}(t)\rangle$ serves as an auxiliary vector. From Eq. (\ref{HS}) we can derive
\begin{eqnarray}
\label{HS1}
H_{S}(t)\!=\!&\Delta(t)|e\rangle\langle e|+\!\left\{\Omega_{S}(t)e^{-i[\beta(t)\!+\!\chi(t)]}|\mu_{2}(0)\rangle\langle e|\!+\!\rm{H.c.}\right\}, \notag \\
\end{eqnarray}
with
\begin{eqnarray}\label{DDpara}
&&\triangle(t)=-\dot{\beta}(t)\left[1+\cos\alpha(t)\right], \quad \Omega_{S}(t)=\frac{1}{2}\sqrt{\left[\dot{\beta}(t)\sin\alpha(t)\right]^{2}
+\dot{\alpha}^{2}(t)},\notag \\
&&\chi(t)=\arctan\left\{\dot{\alpha}(t)\big{/}\left[\dot{\beta}(t)\sin\alpha(t)\right]\right\}, \quad \dot{\zeta}(t)=\frac{1}{2}\dot{\beta}(t)[3+\cos\alpha(t)].
\end{eqnarray}

\begin{figure}[t]
  \centering
  \includegraphics[width=0.7\linewidth]{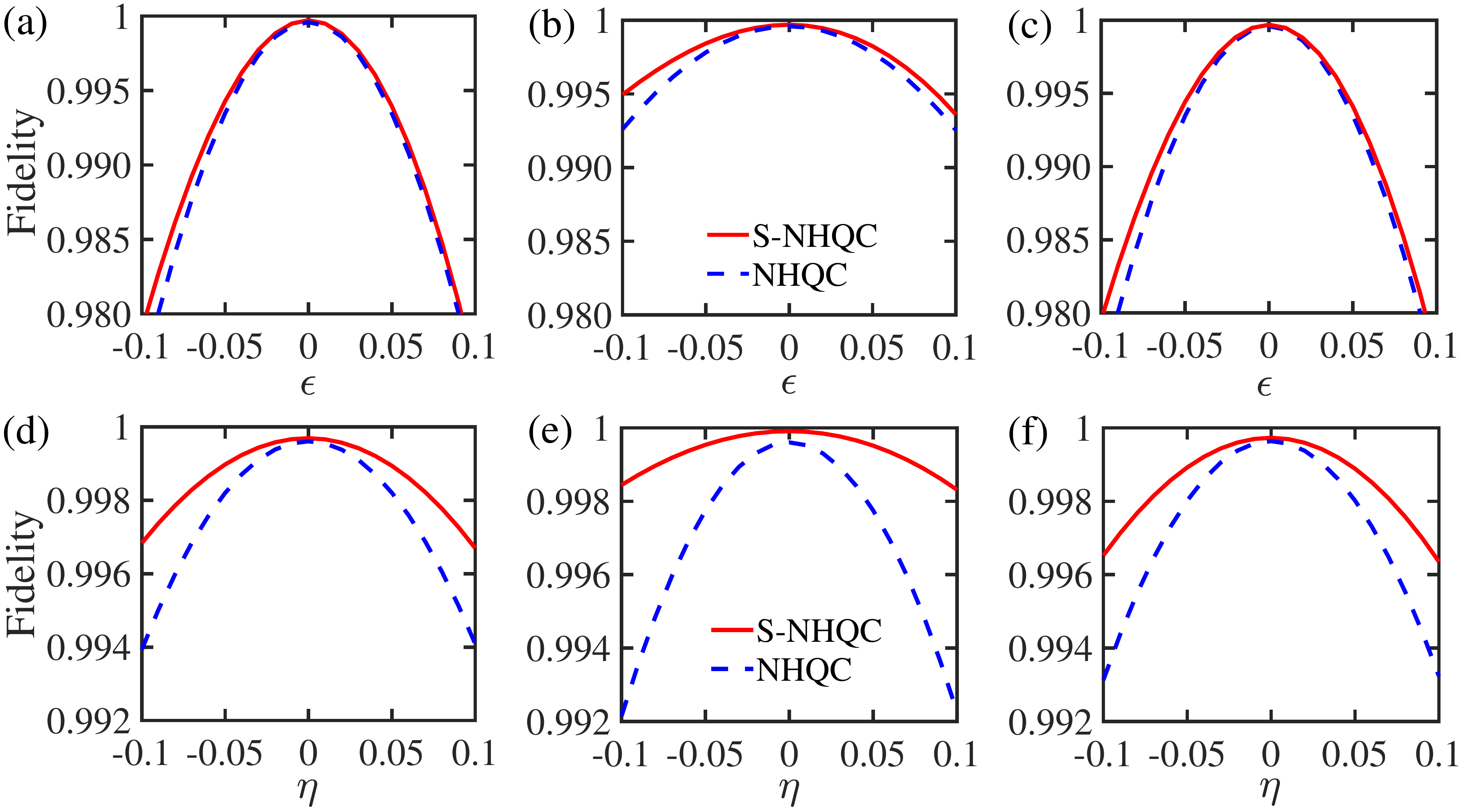}
  \caption{The robustness of the (a) $S$ gate, (b) $T$ gate, and (c) $\sqrt{H}$ gate
to the systematic Rabi error. Robustness of the (d) $S$ gate, (e) $T$ gate, and (f) $\sqrt{H}$ gate to the detuning error. The solid-red and dashed-blue lines denote the results from S-NHQC and NHQC schemes, respectively. Reproduced with permission \cite{liang2022composite}. Copyright 2022, American Physical Society.} \label{shortest-path-robustness}
\end{figure}

To generate the nonadiabatic holonomic gate with the shortest path, we provide a general set of parameter forms of $\alpha(t)$ and $\beta(t)$ that can be used to create a circular path, i.e.,
\begin{eqnarray}\label{circlepara}
\beta(t)=\beta_{0}+\pi\sin^{2}\left(\frac{\pi t}{2\tau}\right), \quad
 \alpha(t)=2\arctan[\ell\sin(\beta(t)-\beta_{0})],
\end{eqnarray}
with $\ell=\sqrt{2\pi \gamma-\gamma^2}/(\pi-\gamma)$, and the geometric phase $\gamma=\int^{\tau}_0\dot{\beta}(t)[1-\cos\alpha(t)]dt/2$. The circular paths of different holonomic gates are shown in Fig.~\ref{shortest-path}(a), it can be observed that the smaller the rotation angle, the shorter the circular evolution path. In addition, as depicted in Fig.~\ref{shortest-path}(b), the smaller the rotation angle, the shorter the evolution time. Benefiting from the shortened evolution path, the robustness of the gates here are also improved compared to the SL-NHQC, as demonstrated in Fig.~\ref{shortest-path-robustness}.

\subsection{NHQC based on shortcuts to adiabaticity}

While NHQC speeds up gate operations compared to adiabatic HQC, it lacks robustness to systematic errors and requires more precise control in experiments. To tackle these problems,  researchers suggested the shortcut to adiabaticity (STA) based NHQC   \cite{guery2019shortcuts, song2016shortcuts,zhang2015fast, liu2017superadiabatic, huang2018shortcut, kang2018nonadiabatic, du2019geometric, kang2020heralded,  liu2020leakage, li2022dynamical, kang2020flexible}.

Shortcut to adiabaticity based on transitionless driving technique is to seek a transitionless Hamiltonian that can drive the system to evolve accurately at any desired rate along the adiabatic channel.  The transitionless driving Hamiltonian can be written as \cite{zhang2015fast}
  \begin{eqnarray}
\label{}
H_{t}=H_{t0}+H_{t1}
=\sum_{n,k}E_n|\varphi_k^n\rangle\langle \varphi_k^n| +i\sum_{n,k,l}(|\dot{\varphi}_k^n\rangle\langle\varphi_k^n|-\langle \varphi_k^n |\dot{\varphi}_l^n\rangle|\varphi_k^n\rangle\langle\varphi_l^n|),
\end{eqnarray}
where $H_{t0}=\sum_{n,k}E_n|\varphi_k^n\rangle\langle \varphi_k^n|$ is the target Hamiltonian, $H_{t1}=i\sum_{n,k,l}(|\dot{\varphi}_k^n\rangle\langle\varphi_k^n|-\langle \varphi_k^n |\dot{\varphi}_l^n\rangle|\varphi_k^n\rangle\langle\varphi_l^n|)$ is the extra Hamiltonian term that enables the dynamical evolution of the system to proceed entirely along an adiabatic channel driven by $H_{t0}$. $|\varphi_k^n\rangle$ are a set of degenerate eigenstates of Hamiltonian $H_{t0}$, with the corresponding eigenvalues $E_n$.

When a generic system of four bare states is considered \cite{zhang2015fast}, in the interaction picture, the Hamiltonian  is $ H_{t0}=\Omega_{t0}|e\rangle(f_{eo}\langle 0|+f_{e1}\langle 1|+f_{e2}\langle 2|)+{\rm H.c.}$, where $f_{ek}$ are the time-dependent control parameters for transitions between $|k\rangle \leftrightarrow |e\rangle$ ($k=0,\ 1,\ 2$). The qubit information is stored in the dark state subspace to prevent any dynamical contribution, and the adiabatic holonomic quantum gate can be obtained through the implementation discussed in \cite{duan2001geometric}. To speed up the adiabatic evolution, an additional Hamiltonian term $H_{t1}$ is added to the target Hamiltonian $H_{t0}$. Specifically, when constructing the phase-shift gate $U_P={\rm exp}(i\gamma_1|1\rangle\langle 1|)$, the Hamiltonian can be rephrased as $H_{t0}=\Omega_t|e\rangle(-\sin(\theta/2) {\rm exp}(i\varphi)\langle 1|+\cos(\theta/2)\langle 2|)+{\rm H.c.}$. The shortcut to this phase-shift gate is realized by adding an extra Hamiltonian term $H_{t1}$. To simplify the form of $H_{t1}$, we divide the evolution process into three steps, resulting in an ``orange slice" path on the parameter sphere. In the first step, the evolution path starts at the North Pole and follows the longitude to the South Pole. We choose $\varphi=0$ and $H_{t1}=\dot{\theta}/2 |2\rangle\langle 1|+{\rm H.c.}$, so $H_t=H_{t0}+H_{t1}$ becomes a $\Delta$-like Hamiltonian. In the second step, we keep $\theta=\pi$  while varying $\varphi$ from 0 to $\varphi_1$, creating a total Hamiltonian that is just a transition between $|e\rangle$ and $|1\rangle$ with detuning $\dot{\varphi}$. Finally, we keep $\varphi_1$ constant and return to the North Pole along the longitude. In this case, the total Hamiltonian is of the same form as the first step. Due to the additional Hamiltonian term $H_{t1}$, the evolution can proceed rapidly along the dark state channel without having to satisfy the adiabatic condition.

The STA-based NHQC has been demonstrated experimentally  \cite{yan2019experimental}   in a three energy levels system with a superconducting qubit in a scalable architecture. In their scheme, all single-qubit holonomic gates can be achieved nonadiabatically through a single-loop evolution. Characterized by QST, the fidelities of the $X$, $H$, and $X(\pi/2)$ gates were found to be $F_X=96.6\pm0.8\%$, $F_H=97.6\pm1\%$, and $F_{X(\pi/2)}=96.4\pm1\%$, respectively. The errors were attributed to higher energy level contributions, decoherence, and control pulse errors.

\section{Enhancing the Robustness of holonomic gates}\label{robustness}

As outlined in the above section, the holonomic gate fidelity can be improved through the time-optimal technique and short-path schemes, but these schemes only slightly increase the gate robustness against error. To enhance the gate robustness, optimization methods such as pulse shaping, composite pulse, and dynamical corrected can be used. However, these schemes come at the cost of increased evolution time, leading to a decrease in maximum achievable fidelity. The NHQC with composite dynamical decoupling scheme, based on the shortest path, can be an alternative that significantly improves both fidelity and robustness.

\subsection{NHQC with encoding}
Quantum computation will inevitably suffer  from  noises, making it difficult to scale up in experiments. To overcome this constraint, error correction and fault tolerance strategies are crucial for achieving scalable quantum computation in the future. One effective approach is the combination of NHQC with  DFS which protects quantum gates from both manipulation errors and decoherence \cite{xu2012nonadiabatic, liang2014nonadiabatic, zhou2015cavity, xue2015universal, pyshkin2016expedited, xu2014universal}. We consider a quantum system interacting with the environment and the total Hamiltonian is
 \begin{eqnarray}
\label{}
H_{SB}=H_S\otimes I_B+I_S\otimes H_B+H_I,
\end{eqnarray}
where $I_S$ and $I_B$ are the identity operator of quantum system and its environment, $H_S$ and $H_B$ represent the Hamiltonian of the system and the environment, respectively. The interaction Hamiltonian between the quantum system and its environment is represented by $H_I=\sum_a S_a\otimes B_a$, with $S_a$ and $B_a$ being the operators for the system and environment.
We suppose that there is an eigenspace $S={\rm Span}\{|\psi_k\rangle\}_{k=1}^n$ spanned by the degenerate eigenstates, and $S_a|\psi_k\rangle=\lambda_a|\psi_k\rangle$, $H_S|\psi_k\rangle\in S$. Therefore, the quantum state starting from $S$ will not be affected by environmental noise, i.e., $S$ is a decoherence-free subspace. Furthermore, if an $L$-dimensional subspace $S^L={\rm Span}\{|\psi_{j}\rangle\}_{j=1}^L$ ($S^L\subset S$) exists and satisfies both the cyclic evolution condition and parallel evolution condition, a nonadiabatic holonomic gate acted on this subspace can be constructed by encoding the qubit within $S^L$.

It was first proposed to combine DFS with NHQC in Ref. \cite{xu2012nonadiabatic}.
The quantum system there consists of $N$ physical qubits interacting with a phase-shifting environment. The system Hamiltonian is
\begin{eqnarray}
\label{}
H_{df}={1 \over 2} \sum_{k<l}^N [J_{kl}^x (\sigma_k^x\sigma_l^x+\sigma_k^y\sigma_l^y) +J_{kl}^y (\sigma_k^x\sigma_l^y-\sigma_k^y\sigma_l^x)],
\end{eqnarray}
where $J_{kl}^x $ and $J_{kl}^y $ are controllable coupling constants between qubits, the first and second terms are $XY$ and Dzialoshinski-Moriya interaction terms, respectively. $\sigma_k^{\alpha} \quad (\alpha=x,y,z)$ is the Pauli operator that operates on the $k$th qubit. The interaction between the system and the environment is described by the Hamiltonian $H_I=(\Sigma_k\sigma_k^z)\otimes B$, where $\Sigma_k\sigma_k^z$ is the collective spin operator for the system, and $B$ is the corresponding operator for the environment. For a system with three physical qubits, the subspace $S=\rm{Span}\{|100\rangle, |010\rangle, |001\rangle\}$ is a three-dimensional DFS. We can choose $S^L=\rm{Span}\{ |010\rangle, |001\rangle\}$ to be the computational space and denote the logical qubits as $|0\rangle_L=|010\rangle$ and  $|1\rangle_L=|001\rangle$. The remaining vector $|100\rangle$ acts as the ancillary qubit. The parameters of Hamiltonian are set as
 \begin{eqnarray}
\label{}
J_{12}^x=-J_{13}^x=J(t)\cos(\phi/2),\quad
J_{12}^y=J_{13}^y=-J(t)\sin(\phi/2).
\end{eqnarray}
 When $\int_0^{\tau}J(t)dt=\pi/\sqrt{2} $ is satisfied, the evolution operator acting on the computational space is a nonadiabatic holonomic gate.
However, this approach requires using two non-commuting one-qubit gates to achieve an arbitrary one-qubit gate, which results in a long evolution time. To shorten the evolution time, researchers combined the use of DFS and a single-loop implementation to construct the nonadiabatic holonomic gate within one loop in various physical systems \cite{xue2016nonadiabatic,hu2016multi, pyshkin2016expedited, sun2016non, zhao2016nonadiabatic, song2016shortcuts, lin2017holonomic, liu2017universal, wang2018single, mousolou2018entangling, ji2019scalable, chen2020robust}, such as superconducting circuit, quantum dots, trapped ions, nitrogen-vacancy  centers, cavity QED, etc.

Aside from the DFS, other quantum error-correcting codes have been designed to protect stored quantum information from noise and decoherence. The binomial code is particularly important in correcting the dominant photon-loss errors in bosonic modes \cite{chen2021fast}. To make the quantum computation solutions more resistant to collective decoherence, one can combine the HQC with passive protection methods such as noiseless subsystems \cite{zhang2014quantum}. Additionally, surface codes are considered promising candidates for achieving large-scale, fault-tolerant quantum computation and have been used to protect holonomic quantum gates \cite{zheng2015fault,zhang2018holonomic,wu2020holonomic}.

\subsection{NHQC with pulse shaping }

It is found that one can design target pulse-shaping NHQC (PS-NHQC) that satisfies both experimental requirements and enhances the robustness of the proposed scheme.   And, the NHQC scheme based on pulse shaping has been successfully demonstrated experimentally \cite{yan2019experimental, ai2020experimental, dong2021experimental,ai2022experimental}.

Pulse shaping is often applied to the Hamiltonian acquired through reverse engineering to enhance its robustness to systematic Rabi error \cite{liu2019plug, kang2019one, li2020fast, li2021multiple}. The Hamiltonian under consideration has parameters that must be determined, in the same form as presented in Eq.~(\ref{sl}), that is,
 \begin{eqnarray}
\label{HP}
H_{ps}(t)=\Omega_{ps}(t)e^{-i\phi_0(t)}|b\rangle\langle e|+\rm{H.c.},
\end{eqnarray}
while the phase $\phi_0(t)$ is time-dependent.
The evolution states that fulfill the Shr\"{o}dinger equation and are spanned by $\{|b\rangle, |e\rangle\}$ can be described by using two time-dependent angles, $\chi(t)$ and $\varphi(t)$, and a time-dependent global phase, i.e.,
\begin{eqnarray}
\label{psip}
|\Psi_{ps}(t)\rangle &=&e^{-i\frac{f(t)}{2}}
\left(
\begin{array}{cc }
e^{-i\frac{\varphi(t)}{2}} \cos[\frac {\chi(t)}{2}]|b\rangle+
 e^{i\frac{\varphi(t)}{2}}\sin[\frac {\chi(t)} {2}]|e\rangle
\end{array}
\right).
\end{eqnarray}
By solving the Shr\"{o}dinger equation $i|\dot{\Psi}_{ps}(t)\rangle= H_{ps}(t)|\Psi_{ps}(t)\rangle$, the relationships between the parameters can be obtained as
\begin{eqnarray}
\label{PSpara}
\dot{f}(t)&=&-\dot{\varphi}(t)/\cos\chi(t), \quad
\dot{\chi}(t)=-2\Omega_{ps}(t)\sin[\varphi(t)+\phi_0(t)],\notag\\
\dot{\varphi}(t)&=&-2\Omega_{ps}(t)\cot\chi(t)\cos[\varphi(t)+\phi_0(t)].
\end{eqnarray}
In particular, when $\chi(t)$ satisfies the cyclic condition $\chi(0)=\chi(\tau)=0$, the evolution state will accumulate a phase factor that includes both the geometric and the dynamical components after a periodic evolution. Thus, the holonomic gate can be acquired by eliminating the dynamical phase. Furthermore, the pulse shape of $\Omega_{ps}(t)$ is determined by the free parameters $f(t)$ and $\chi(t)$, so we can design a special pulse shape to suppress the influence of noise or error by choosing appropriate parameters.

For the  Rabi error, i.e., $\Omega_{ps}(t)\rightarrow(1+\epsilon)\Omega_{ps}(t)$, the excitation profile at the end of the first interval is
\begin{eqnarray}
\label{}
P_{\epsilon}=\left|\langle\Psi_{ps}(\tau/2)|\Psi^{\epsilon}_{ps}(\tau/2)\rangle\right|^2
\simeq 1-{\epsilon}^2\left |\int^{\tau/2}_0e^{-if}\dot{\chi}\sin^2\chi dt\right|^2,
\end{eqnarray}
 with $\tau$ being the total time of the evolution. This shows that the gate is resistant to first-order of Rabi error. To make it even more robust, we establish $\chi(t)=\pi\sin^2(\pi t/\tau)$, $f(t)=\varsigma[2\chi-\sin(2\chi)]$, thus
\begin{eqnarray}
\label{}
P_{\epsilon}&\simeq1-{\epsilon}^2\sin^2(\varsigma\pi)/(2\varsigma)^2.
\end{eqnarray}
Hence, for a positive integer $\varsigma$, we can achieve $P_{\epsilon}=1$, which eliminates the influence of the second-order systematic Rabi error completely. However, under the limit of a certain maximum of pulse amplitude, $\varsigma\geq1$ corresponds to a very long evolution time, which will lead to a significant effect of decoherence and reduce the gate-fidelity. Therefore, we need to determine the optimal value of $\varsigma$ to strike a balance between fidelity and robustness.

\begin{figure}[t]
  \centering
  \includegraphics[width=0.5\linewidth]{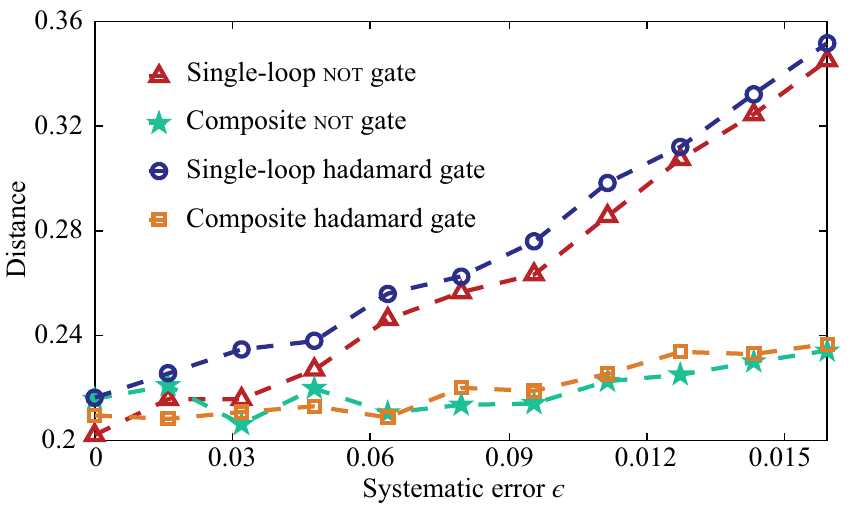}
  \caption{Experimental gate distances to the Rabi error $\epsilon$ for single-qubit holonomic $NOT$ and $Hadamard$ gates in the single-loop scheme and in the composite scheme with $N = 2$. Adapted with permission \cite{zhu2019single}. Copyright 2019, American Physical Society.} \label{figure7-multiloop}
\end{figure}

The optimal control of this pulse shaping has been confirmed in trapped ion system \cite{ai2020experimental}, where they set $\varsigma=1/5$. The performance of the implemented holonomic single-qubit quantum gates was characterized using the interleaved RB method, and the resulting fidelities were $F_X=99.10\%$, $F_H=98.90\%$, $F_T=99.20\%$, and $F_S=99.10\%$, respectively. The study also demonstrated the robustness of the optimized gates to Rabi error. The pulse shaping NHQC exhibited much greater robustness compared to the conventional SL-NHQC scheme and even surpassed the dynamical scheme in this aspect.

\subsection{ NHQC with composite pulse}

The composite NHQC (C-NHQC) is a simple and effective method to suppress the influence of Rabi error, especially for large-angle rotation operations \cite{xu2017composite,zhu2019single}. The central idea of the composite scheme is to take $U(\gamma/N, \theta, \phi)$ as an elementary gate with rotation angle of $\gamma/N$, and then the target gate $U(\gamma, \theta, \phi)$ is achieved by consecutively applying $N$ of these elementary gates, causing the accumulated geometric phase to be $\gamma$, i.e., $[U(\gamma/N, \theta, \phi)]^N=U(\gamma, \theta, \phi)$. Here $\gamma$ is the rotation angle, and $\theta$ and $\phi$ are used to determine the rotation axis.

It is experimentally verified that \cite{zhu2019single}  the C-NHQC scheme is indeed more robust compared to the single-loop scheme when it comes to Rabi error, as illustrated in Fig.~\ref{figure7-multiloop}. They construct the holonomic NOT gate and Hadamard gate with two loops ($N=2$), and the corresponding evolution operator were $NOT=[U(\pi/2, \pi/2, 0)]^2$ and $H=[U(\pi/2, \pi/4, 0)]^2$, respectively. The abnormal behaviors of NOT gates within a small systematic Rabi error range mainly come from the imperfect state preparation and measurement which accounts for the dominant error when the Rabi error is small.

Even though the composite pulse enhances robustness against Rabi frequency error, it also prolongs the evolution time. This is because the evolution time for each cycle remains the same, regardless of whether the rotation angle is large or small. For instance, when $N$ is 2, $U(\gamma, \theta, \phi)=[U(\frac{\gamma}{2}, \theta, \phi)]^2$, the composite scheme requires twice as much evolution time compared to the single-loop scheme. This makes the system more vulnerable to environment-related decoherence. This is why the composite scheme has lower fidelity compared to the single-loop scheme when the decoherence is taken into account.

\begin{figure}[h]
  \centering
  \includegraphics[width=0.95\linewidth]{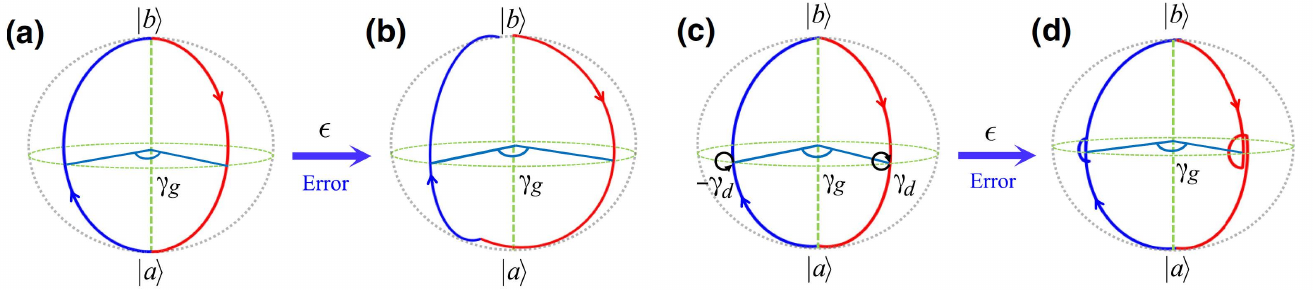}
  \caption{Schematic diagram of the evolution paths for holonomic quantum gate. The evolution path of the single-loop NHQC scheme is shown in (a) for the ideal case and in (b) for the case with the Rabi error. The evolution path of the DC-NHQC scheme is shown in (c) for the ideal case and in (d) for the case with Rabi error. As demonstrated in (d), the Rabi error can be effectively corrected by the dynamic correction. Reproduced  with permission \cite{li2021dynamically}. Copyright 2021, American Physical Society.} \label{figure11-DCG}
\end{figure}

\subsection{NHQC with dynamically correction}

With the same purpose as pulse shaping and composite pulse optimization, dynamically corrected NHQC (DC-NHQC) \cite{li2021dynamically,liu2021super,he2021robust} is also an optimization method that improves the gate robustness.   The idea of DC-NHQC is illustrated  in Fig.~\ref{figure11-DCG}. Figure~\ref{figure11-DCG}(a) illustrates the evolution path of the conventional single-loop NHQC without interference from errors. In the absence of noise, the path of the evolution state in the parameter space travels from the North Pole to the South Pole and then returns to the North Pole after passing through a phase change at the South Pole. However, when there is systematic Rabi error present, as expressed by $H^{\epsilon}_{dc}(t)=(1+\epsilon)H_{sl}(t)$, where $\epsilon$ is the error fraction and $H_{sl}(t)$ is the same form as Eq.~(\ref{sl}), the evolution path is no longer a perfect orange-slice path, as shown in Fig. \ref{figure11-DCG}(b). In this case, when the systematic Rabi error is small, i.e., $\epsilon\ll1$, the gate fidelity drops from perfect unity to \cite{li2021dynamically}
\begin{eqnarray} \label{}
F = 
1- {1 \over 2} \left({\epsilon \pi \over 2}\right)^2 (1-\cos\gamma_g),
\end{eqnarray}
where 
$\gamma_g$ is the pure geometric phase. Thus,  due to the Rabi error, and the trajectory of the evolution can not precisely return to the North Pole in the  single-loop NHQC scheme.

To suppress the influence of systematic Rabi error through dynamical correction, we interpolate two dynamical processes at the halfway point of each half-evolution path. The sum of dynamical phases in these two processes is zero. The first dynamical process is governed by the Hamiltonian
\begin{eqnarray}
\label{dcinsert1}
\mathcal{H}_1(t)=(1+\epsilon)\Omega_{dc}(t)e^{-i(\phi_0+\pi/2)}|b\rangle\langle e|+{\rm H.c.},
\end{eqnarray}
which is inserted at time $T/4$ with $\int^{3T/4}_{T/4}\Omega_{dc}(t)dt=\pi/2$. The Hamiltonian
\begin{eqnarray}
\label{dcinsert2}
\mathcal{H}_2(t)=(1+\epsilon)\Omega_{dc}(t)e^{-i(\phi_0-\gamma_g-\pi/2)}|b\rangle\langle e|+{\rm H.c.}
\end{eqnarray}
with $\int^{7T/4}_{5T/4}\Omega_{dc}(t)dt=\pi/2$ is inserted at $5T/4$ to drive the second dynamical process.
The trajectory on the Bloch sphere is depicted in Fig.~\ref{figure11-DCG}(c) without Rabi error and Fig.~\ref{figure11-DCG}(d) with Rabi error, respectively. In this case, despite the presence of Rabi error in all the processes, the trajectory can still approximately return back to the North Pole. The gate fidelity is calculated to be
\begin{eqnarray}
\label{}
F_c = 
1-{1 \over 2} \left({\epsilon \pi \over 2}\right)^4(1-\cos\gamma_g),
\end{eqnarray}
indicating that the dynamical correction can suppress the error to the fourth-order perturbation, which greatly enhances the robustness of the Rabi error.

\begin{figure}[!t]
  \centering
\includegraphics[width=0.8\linewidth]{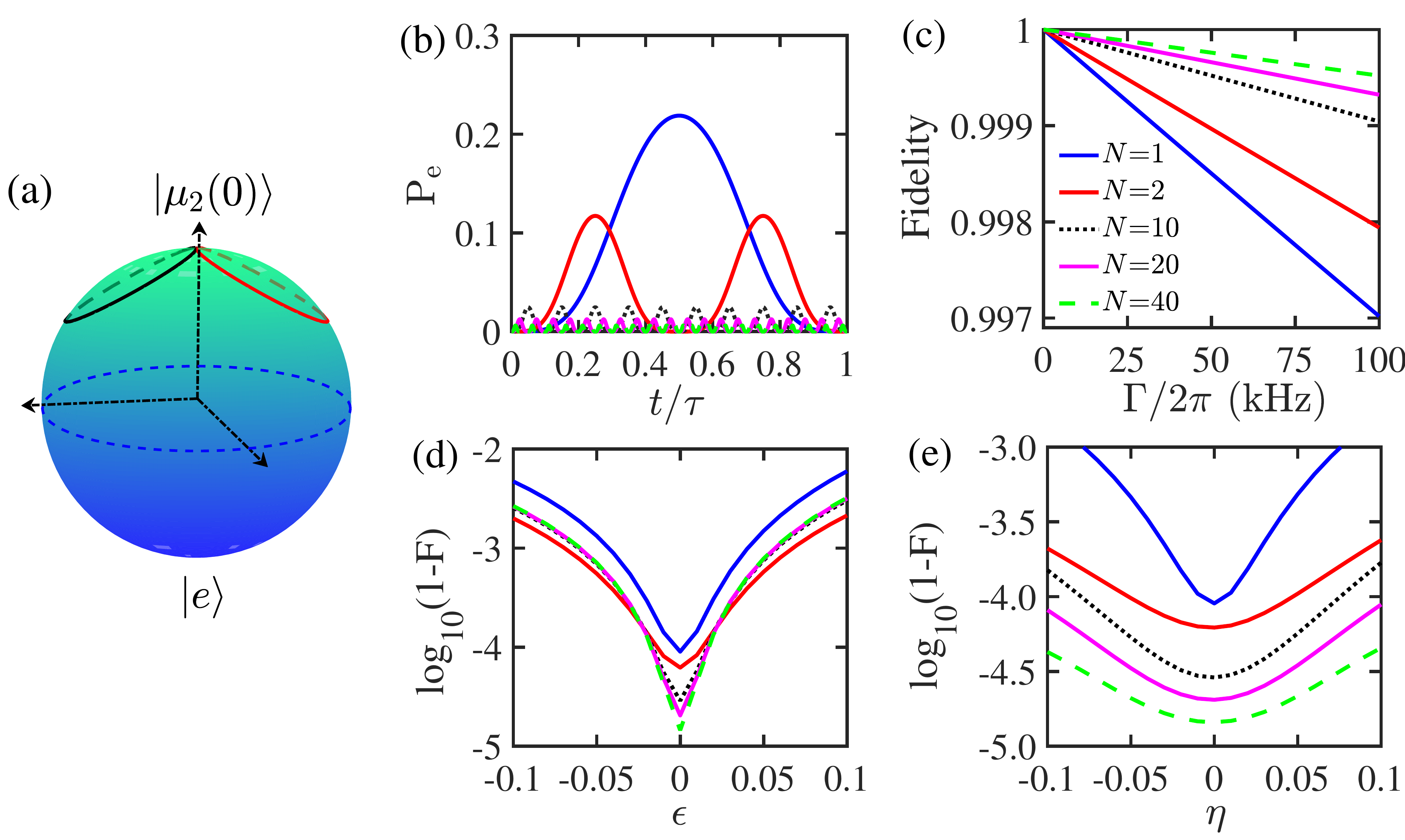}
\caption{(a) The symmetrical evolution path of the $T$ gate obtained by using the simplest composite dynamical decoupling pulse with $N=2$. The performance of the $T$ gate optimized with different  pulse sequences, where (b) represents the population of the excited state, and (c) depicts the gate fidelity as a function of decoherence. The comparison with the un-optimized S-NHQC scheme (blue lines), where (d) represents the gate infidelity with respect to systematic Rabi error, and (e) represents the gate infidelity with respect to detuning error. Reproduced with permission \cite{liang2022composite}. Copyright 2022, American Physical Society.} \label{CDD_estate_robust}
\end{figure}

\subsection{NHQC with composite dynamical decoupling}

The environment-induced decoherence is one of the obstacles to the realization of high-fidelity and robust quantum gates. Dynamical decoupling \cite{viola1999dynamical} can effectively reduce the impact of decoherence by reversing the evolution of the quantum system through the use of control pulses at specific times, which can be applied \cite{xu2018path, zhao2021dynamical} to the NHQC. The physical system under consideration consists of $N$ qubits, and the interaction between the system and the environment is determined by the Hamiltonian $H_I=\sum_{k, a}\sigma_k^a\otimes B_k^a$, where $B_k^a$ is the environment operator corresponding to the Pauli operator $\sigma_k^a$ ($a=x, y, z$) in the $k$th qubit. The impact of environment-induced decoherence can be suppressed through the use of dynamical decoupling with a decoupling sequence of the decoherence-suppressed being $ \{\otimes_{k=1}^NI_k, \ \otimes_{k=1}^N\sigma_k^x, \ \otimes_{k=1}^N\sigma_k^y,\  \otimes_{k=1}^N\sigma_k^z\}$. However, this decoupling sequence may interfere with the evolution of the quantum system. To make the decoupling sequence compatible with the system evolution, we chose a Hamiltonian of the form $\sigma_k^a\sigma_j^a$ ($k, j=1,2,..., N$) to implement NHQC.

However, the traditional dynamical decoupling method requires the addition of fast and strong decoupling sequences, which can lead to control errors. To address this issue, we proposed a   composite dynamical decoupling NHQC (CDD-NHQC) scheme\cite{liang2022composite}, which does not require the addition of external field control. According to the cyclic and parallel transport conditions, the Hamiltonian of the single-qubit holonomic gate obtained through inverse engineering is the same as in Eq.~(\ref{HS}). To decouple the gate implementation from the environment, the system evolution is divided into $N$ segments, resulting in $U(\gamma) = U_N(\gamma/N) \cdots U_2(\gamma/N)U_1(\gamma/N)$. By setting $\beta_{0}$ in all even-numbered pulse sequences to $\beta_{0}+\pi$, the evolution paths are symmetrical about the North Pole on the Bloch sphere, and thus effectively suppressing decoherence.

Figure~\ref{CDD_estate_robust}(a) shows the evolution paths of the $T$ gate when $N=2$, and Fig.~\ref{CDD_estate_robust}(b) demonstrates that the population of the excited state $|e\rangle$ decreases as the number of pulse sequences $N$ increases, reducing decoherence, see Fig.~\ref{CDD_estate_robust}(c). Additionally, the CDD-NHQC scheme can enhance the robustness of systematic Rabi error and detuning error. As shown in Figs.~\ref{CDD_estate_robust}(d) and (e), the robustness of the $T$ gate to detuning error becomes stronger with increasing $N$, while the robustness to Rabi error reaches saturation at $N=2$.

\section{Discussions and conclusions}

\subsection{ Gate performance for different optimal strategies}

As discussed in the above, various optimized schemes are devoted to improving the gate-fidelity and enhancing the robustness against errors. Here we compare the robustness of the $S$ gate for recently proposed optimization schemes, using an ideal three-level $\Lambda$ system, the detailed parameters for each scheme are shown in Table \ref{tab1}. We introduce the systematic Rabi error, i.e., $\Omega\rightarrow(1+\epsilon)\Omega$, and the detuning error $\eta \bar{\Omega}|e\rangle\langle e|$, where $\epsilon$ and $\eta$ are Rabi error rate and detuning error rate, and $\bar{\Omega} $ is the average value of the Rabi frequency.
The performance of different  schemes can be evaluated by the Markovian master equation in the Lindblad form
\begin{eqnarray}
\label{EqMaster}
\dot\rho&=&-i[H_i(t), \rho]+\frac {1} {2}\sum_{j=-,z}\Gamma_{j}L(\sigma_{j}),
\end{eqnarray}
where $\rho$ is the density matrix of the quantum system, $H_i(t)$ is the corresponding Hamiltonian in various schemes, and $L(A)=2A\rho A-A^{\dag}A\rho-\rho A^{\dag}A $ is the Lindbladian operator with $\sigma_-=|0\rangle\langle e|+|1\rangle\langle e|$ and $\sigma_z=|e\rangle\langle e|-|1\rangle\langle1|-|0\rangle\langle0|$, $\Gamma_-$ and $\Gamma_z$ represent the decay and dephasing rates, respectively.

\begin{table}[!t]
\footnotesize
\caption{Parameters  for quantum gates with different optimal strategies.}
\label{tab1}
\tabcolsep 3pt %
\begin{tabular*}{\textwidth}{llrrrllc}
\toprule

Scheme  &      Hamiltonian & \ & \ &\  &          Pulse shape &          Phase &  Pulse area  \\
 \hline
&&&&\\
  SL-NHQC  &                  Eq. (\ref{sl})& &&&        $\Omega_{sl}=\bar{\Omega}$&     $  \left\{
                                                                                             \begin{array}{ ll}

                                                                                              \phi_0=0, & t\in[0,T/2],   \\

                                                                                               \phi_0=-\pi/2, & t\in[T/2,T].

                                                                                              \end{array}\right.$                                      &     $\pi$\\

 &&&&\\

\hline
&&&&\\
PS-NHQC  &                    Eq. (\ref{HP})&  &&&              $  \left\{

                                                       \begin{array}{l}

                                 \Omega_{ps}(t)=\sqrt{\Omega^2_{R}+\Omega^2_{I}},   \\

                                       \Omega_R =\cos\varphi\sin\chi \dot{f}-\sin\varphi \dot{\chi},\\

                                        \Omega_I=\sin\varphi\sin\chi \dot{f}+\cos\varphi \dot{\chi},\\

                                        \chi=\pi\sin^2[\pi t/(2\tau)]                              \\

                                        f=[2\chi-\sin(2\chi)],                                    \\

                                      \int^{T}_0\Omega_{ps}(t)dt/T=\bar{\Omega}
                                                      \end{array}\right.$   &              $  \left\{

                                                                                             \begin{array}{ l}

                                                                                             \dot{\varphi}=-\dot{f}\cos\chi, \\

                                                                                              \varphi(0)=-\pi/2, \\

                                                                                               \varphi(T/2)=-\pi ,\\

                                                                                                \tan\phi_0=\Omega_I/\Omega_R,

                                                                                                \end{array}\right.$                                 &  $2.16\pi$\\

&&&&\\

\hline
&&&&\\
C-NHQC &                    Eq. (\ref{sl})& &&&        $\Omega_C=\bar{\Omega}$ &                        $  \left\{

                                                                                             \begin{array}{ ll}

                                                                                              \phi_0=0, & t\in[0,T/2],   \\

                                                                                               \phi_0=-\pi/4, & t\in[T/2,T],\\

                                                                                                \phi_0=0, & t\in[T,3T/2],   \\

                                                                                                 \phi_0=-\pi/4, & t\in[3T/2,2T].

                                                                                                \end{array}\right.$                                 &       $2\pi$ \\

&&&&\\

 \hline
&&&&\\
DC-NHQC   &    Eqs. (\ref{sl}), (\ref{dcinsert1}), (\ref{dcinsert2}) & &&&  $\Omega_{dc}=\bar{\Omega}$   &   $  \left\{

                                                                                             \begin{array}{ ll}

                                                                                              \phi_0=0, & t\in[0,T/4],   \\

                                                                                               \phi_0=\pi/2, & t\in[T/4,3T/4],\\

                                                                                                \phi_0=0, & t\in[3T/4,T],   \\

                                                                                                \phi_0=\pi/2, & t\in[T,5T/4],\\

                                                                                                \phi_0=-\pi, & t\in[5T/4,7T/4],\\

                                                                                                 \phi_0=\pi/2, & t\in[7T/4,2T].

                                                                                                \end{array}\right.$                                &      $2\pi$\\

&&&&\\

 \hline
&&&&\\
TO-UNHQC  &          Eq. (\ref{HTO})& &&&                     $\Omega_T=\bar{\Omega}$   &            $ \phi=\pi t/\tau$                    &      $0.43\pi$ \\

&&&&\\

\hline
&&&&\\

S-NHQC  &           Eq. (\ref{HS})&&&&  $  \left\{

                           \begin{array}{ l}

                          \Omega_S(t)=\sqrt{[\dot{\beta}\sin\alpha]^2+\dot{\alpha}^2}/2   \\

                           \int^{T}_0\Omega_S (t)dt/T=\bar{\Omega}

                              \end{array}\right.$  &                                    $  \left\{

                                                                                             \begin{array}{l}

                                                                                              \beta= \pi\sin^2(\pi t/2/\tau),   \\

                                                                                               \alpha=2\arctan[l\sin\beta-\beta_0],\\

                                                                                                l=\sqrt{2\pi \gamma-\gamma^2}/(\pi-\gamma),   \\

                                                                                               \chi=\arctan\{\dot{\alpha}/[\dot{\beta}\sin\alpha]\},\\

                                                                                               \gamma=\pi/2,

                                                                                                \end{array}\right.$   &                                          $0.87\pi$\\

&&&&\\

 \hline
&&&&\\

CDD-NHQC  &            Eq. (\ref{HS})&&&& $  \left\{

                           \begin{array}{ l}

                          \Omega_{CD}(t)=\sqrt{[\dot{\beta}\sin\alpha]^2+\dot{\alpha}^2}/2   \\

                           \int^{T}_0\Omega_{CD} (t)dt/T=\bar{\Omega}

                              \end{array}\right.$&                                             $  \left\{

                                                                                             \begin{array}{l}

                                                                                              \beta=\beta_0+ \pi\sin^2(\pi t/2/\tau),   \\

                                                                                               \alpha=2\arctan[l\sin\beta-\beta_0],\\

                                                                                                l=\sqrt{2\pi \gamma-\gamma^2}/(\pi-\gamma),   \\

                                                                                               \chi=\arctan\{\dot{\alpha}/[\dot{\beta}\sin\alpha]\},\\

                                                                                               \beta_0=0, t\in[0,T],\\

                                                                                               \beta_0=\pi, t\in[T,2T],\\

                                                                                               \gamma=\pi/4,

                                                                                                \end{array}\right.$              &                             $1.32\pi$\\

&&&&\\
      \bottomrule
\end{tabular*}
\end{table}

\begin{figure}[!t]
  \centering
\includegraphics[width=0.95\linewidth]{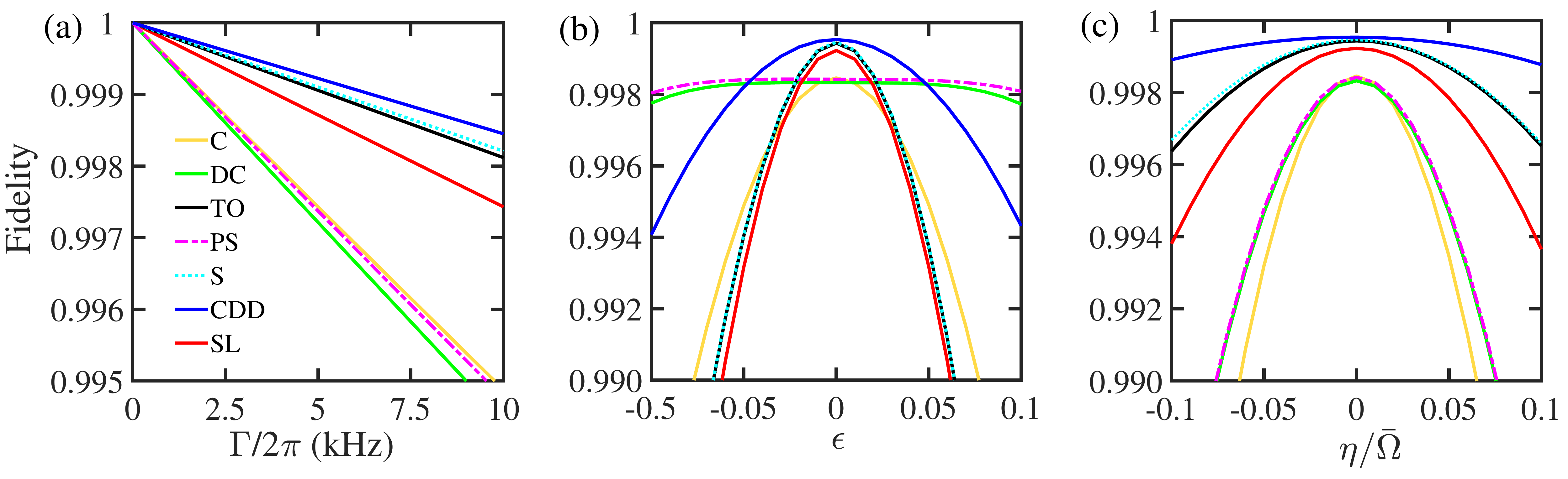}
\caption{ The performance comparison of $S$ gate of different optimal NHQC schemes with $\bar{\Omega}=2\pi\times10 \ {\rm MHz}$. (a) The gate-fidelity as a function of decoherence. The robustness of $S$ gate with respect to (b) Rabi error and (c) detuning error. The decoherence rates are chosen to be $\Gamma_-=\Gamma_z=2\pi\times3$ kHz in (b) and (c).} \label{Fig13}
\end{figure}

As illustrated in Fig.~\ref{Fig13}(a), the TO-UNHQC scheme (black line) performs better in terms of resisting decoherence since it achieves gate operation in the shortest time. However, the S-NHQC (dotted cyan line) and CDD-NHQC (blue line) schemes are more resilient to decoherence, with CDD-NHQC showing the best results, despite the fact that they require longer evolution times than TO-UNHQC. We believe that the ability of S-NHQC and CDD-NHQC schemes to resist decoherence is due to the lower population of excited states compared to the other schemes, as shown in Fig. \ref{shortest-path} and Fig. \ref{CDD_estate_robust}(a). Although the CDD-NHQC method has a long evolution time, it exhibits strong resistance against environment-induced decoherence by decoupling its dynamics from the environment. For systems where decoherence is the main source of error, we can improve fidelity by increasing the number of dynamical decoupling pulse sequences. Conversely, the C-NHQC, PS-NHQC, and DC-NHQC optimization methods are susceptible to decoherence because of their longer evolution time.

Figure~\ref{Fig13}(b) depicts a comparison of the robustness of various optimization schemes against Rabi error. Among these schemes, the PS-NHQC (green line) and DC-NHQC (magenta dot-dashed line) focus on enhancing the robustness of the systematic Rabi error. As anticipated, they exhibit outstanding resistance to Rabi error. The C-NHQC (yellow line) is also an optimization scheme for fortifying against Rabi error, but its effectiveness is not as potent as that of PS-NHQC and DC-NHQC. Nonetheless, it is crucial to note that decoherence is an inevitable factor in real physical systems, which causes the robustness curves of PS-NHQC, DC-NHQC, and C-NHQC to decline, particularly for the DC-NHQC scheme. This implies that high fidelity is compromised in exchange for increased robustness against systematic Rabi error. Achieving satisfactory results necessitates considering the trade-off between various factors in a physical implementation. In general, the CDD-NHQC optimization is a sound choice as it balances high fidelity and robustness. Additionally, it is evident from Fig.~\ref{Fig13}(b) that irrespective of the optimization technique employed, resistance to systematic Rabi error is improved to some degree in comparison to the traditional SL-NHQC scheme (red line).

The robustness to detuning error is demonstrated in Fig.~\ref{Fig13}(c). The CDD-NHQC optimization surpasses other solutions and is the optimal option for mitigating detuning error. For physical systems where the detuning error is the main source of errors, dynamical decoupling pulse sequences can also be increased to enhance robustness while enhancing maximum fidelity. Except for the CDD-NHQC, S-NHQC, and TO-UNHQC optimization methods, the robustness of other optimization methods to detuning error is inferior to the traditional SL-NHQC method.
As different physical systems have distinct primary sources of error, it is preferable to select appropriate optimization procedures based on specific requirements.

\subsection{Conclusion}
In the theoretical design of NHQC,  researchers simplified the original two-loop  scheme to a single-shot  scheme, decreases  the exposure time of  holonomic gates to decoherence. Later, researchers replaced the single-shot solution, which necessitated non-resonant laser pulses, with a single-loop multiple-pulse solution to overcome the limitations of square pulses and limited-rotation-angle. The traditional NHQC, which eliminates the dynamical phase, was subsequently extended to unconventional NHQC, in which the dynamical phase is retained but proportional to the geometric phase. Later on, the cyclic evolution condition was relaxed, and noncyclic NHQC emerged. More recently, NHQC has been developed beyond the three-level setting.
In short, NHQC represents a promising direction for the development of quantum computation and has the potential to offer significant advantages over conventional methods.

\subsection{Perspectives}
While much work remains to be done to fully realize the potential of NHQC, the progress that has been made so far is very encouraging and holds great promise for the future.
Since it was first discovered, HQC has experienced remarkable growth. It has evolved from the early adiabatic case to NHQC, and then to the current stage of optimized NHQC. NHQC inherits geometric characteristics while combining speed with universality, and becomes a promising quantum control solution that has been demonstrated on various  physical platforms. However, quantum computation based on geometric manipulation requires longer computing time and a more complex physical implementation process than traditional dynamical schemes, making it challenging for NHQC to surpass the dynamical scheme and hindering its wide adoption. Therefore,  the development of NHQC schemes that are comparable to the dynamical scheme is crucial for making NHQC widely applicable.

The objective of NHQC optimization is to achieve high-fidelity and robust quantum gates. Certain optimization schemes focus on speeding up the evolution and enhancing fidelity, such as TO-UNHQC and S-NHQC. Others aim to improve robustness to Rabi error, such as PS-NHQC, C-NHQC, and DC-NHQC.
There are also optimization schemes that improve both fidelity and robustness, such as CDD-NHQC. However, the source of errors in quantum systems is complex, and NHQC is susceptible to errors that can impact both the evolutionary path and the path integral. Therefore, one of the most concerning issues in this field is how to combine NHQC with other noise resistance schemes and error correction technologies in quantum information processing to develop a scheme that can resist a variety of errors.

Current experimental research is primarily focused on superconducting circuits, nitrogen-vacancy centers, NMR, and trapped ion systems. Looking ahead, other physical systems such as nanophotonics and Rydberg atoms may also become relevant. While experimental fidelities of single-qubit holonomic gates can exceed 0.99, the fidelities of two-qubit gates remain low, which is insufficient for realizing large-scale quantum computation. The primary reason is due to crosstalk between qubits and leakage from computational subspace to non-computational subspace. Additionally, energy relaxation and dephasing of qubits can also contribute to infidelity. Thus, the most significant challenge facing NHQC is to achieve high-fidelity two-qubit holonomic gates. This is also an urgent problem faced by both geometric quantum computation and dynamical quantum computation.

\bigskip

\noindent {\bf \large Acknowledgements}\\
This work was supported by the National Natural Science Foundation of China (Grant No. 12275090), the Guangdong Provincial Key Laboratory (Grant No. 2020B1212060066), and the Innovation Program for Quantum Science and Technology (Grant No. 2021ZD0302300).





\end{document}